\documentclass[a4paper,11pt]{article}
\usepackage{jheppub} 
\usepackage{lineno}

\usepackage[utf8]{inputenc}
\usepackage{amsmath}
\usepackage{amssymb}
\usepackage{amsfonts}
\usepackage{amsthm}
\usepackage{xcolor}
\usepackage{hyperref}
\usepackage{rotating}
\usepackage{array}
\usepackage{appendix}
\usepackage{bbold}
\usepackage{textcomp}

\usepackage{tikz}
\usetikzlibrary{arrows,chains,matrix,positioning,scopes}

\theoremstyle{definition}

\def\sc{\textit{sc}}

\title{Spinning Partial Waves for Scattering Amplitudes in $d$ Dimensions}
\author{Ilija Buri\'c,}
\author{Francesco Russo}
\author{and Alessandro Vichi}

\affiliation{Department of Physics, University of Pisa and INFN, \\Largo Pontecorvo 3, I-56127 Pisa, Italy}

\emailAdd{ilija.buric@df.unipi.it}
\emailAdd{francesco.russo@phd.unipi.it}
\emailAdd{alessandro.vichi@unipi.it}

\abstract{
Partial wave decomposition is one of the main tools within the modern S-matrix studies. We present a method to compute partial waves for $2\to2$ scattering of spinning particles in arbitrary spacetime dimension. We identify partial waves as matrix elements of the rotation group with definite covariance properties under a subgroup. This allows to use a variety of techniques from harmonic analysis in order to construct a novel algebra of weight-shifting operators. All spinning partial waves are generated by the action of these operators on a set of known scalar seeds. The text is accompanied by a {\it Mathematica} notebook to automatically generate partial waves. These results pave the way to a systematic studies of spinning S-matrix bootstrap and positivity bounds.
}

\begin{document}

\maketitle

\section{Introduction}
In recent years, there has been a resurgence of interest in the study of S-matrix and scattering amplitudes in quantum field theory using a bootstrap approach. This approach aims to construct amplitudes solely by their analytical properties, that express the physical requirements of unitarity and causality \cite{eden2002analytic}. These ideas have been given a fresh look in recent years by combining them with numerical bootstrap methods, \cite{Paulos:2016fap,Paulos:2016but,Paulos:2017fhb}. One of the main targets of this approach has been the scattering of the lightest strongly interacting particle: the pion \cite{Guerrieri:2018uew,Guerrieri:2020bto}. Moreover, there have been recent bootstrap studies of spinning particles, such as fermions \cite{Hebbar:2020ukp} and photons \cite{Haring:2022sdp}.

\smallskip

Besides the non-perturbative S-matrix approach, many recent works have studied constraints on weakly coupled effective field theories (EFTs) of gravity, photons or matter using dispersive arguments. It was realised some time ago that not all choices of EFTs are consistent with a well-behaved UV completion:  unitarity and causality imply positivity bounds on the Wilson coefficients parametrising the EFT action, \cite{Pham:1985cr, Ananthanarayan:1994hf, Adams:2006sv}. Recently, the methods for extracting constraints on EFTs from these basic requirements have been given a more systematic foundation \mbox{\cite{Arkani-Hamed:2020blm,Bellazzini:2020cot,deRham:2017avq, Tolley:2020gtv, Caron-Huot:2020cmc,Sinha:2020win, Trott:2020ebl}}. This has led to a number of important outcomes, including a demonstration that S-matrix consistency implies two-sided bounds on ratios of EFT coefficients, essentially ``proving'' the intuition of dimensional analysis above, as well as a precise numerical recipe for obtaining optimal bounds  \cite{Zhang:2021eeo,Du:2021byy,Davighi:2021osh,Chowdhury:2021ynh,Caron-Huot:2021rmr}. This technique has recently also been applied to pion scattering \cite{Albert:2022oes,Fernandez:2022kzi} but also photons \cite{Henriksson:2021ymi,Henriksson:2022oeu} and gravitons \cite{Bern:2021ppb,Caron-Huot:2021enk,Caron-Huot:2022ugt}.

\smallskip

Both in perturbative and non-perturbative setups, partial wave decomposition has proven to be an essential tool. By decomposing the total scattering amplitude into partial waves, it becomes possible to analyse the scattering process in terms of different angular momentum contributions. For $2\to 2$ scattering of scalars in general dimension $d$, the decomposition takes the form
\begin{equation}\label{scalar-pw-decomposition}
     S(s,t) = 1 + i \sum_{J=0}^{\infty} n^{(d)}_J a_J(s) C_J^{\left(\frac{d-3}{2}\right)}(\cos \theta) \ . 
\end{equation}
Here, $J$ denotes the spin of the exchanged states, $a_J(s)$ is a dynamical function of the Mandelstam variable $s$ and partial waves $C_J^{\left(\frac{d-3}{2}\right)}(\cos \theta)$ are Gegenbauer polynomials of the (cosine of the) scattering angle $\theta$. The convergence of this expansion has been proven to be valid inside the Lehmann-Martin ellipse, see \cite{Correia:2020xtr} for a recent review. \footnote{Note that our terminology differs from some of the literature that uses 'partial wave' to refer to the dynamical function $a_J(s)$.} For particles with spin in $d=4$, the partial waves are given by Wigner-$d$ functions, \cite{wigner1931gruppentheorie}. In certain cases in higher dimensions, these functions have been computed in the context of graviton scattering in \cite{Caron-Huot:2022jli}. The motivation to go beyond four dimensions comes from the well-known infrared divergences appearing in four-dimensional scattering amplitudes involving massless particles \cite{Weinberg:1965nx}.

\smallskip

More in general, partial waves represent an essential building block for scattering amplitudes, very much like conformal blocks are the building blocks of conformal field theories (CFTs). In both cases this allows for a simplified representation of the system, as only a subset of the waves may significantly contribute to the process. It also offers a convenient expansion to parameterise the most general amplitude/correlation function. After the renaissance of the conformal bootstrap \cite{Rattazzi:2008pe,Rychkov:2009ij}, many works have explored the theory of conformal blocks, their mathematical structure and properties, leading to interesting insights and efficient and complementary techniques to compute them (see \cite{Poland:2018epd} for a review).

\smallskip

In order to develop further the S-matrix program, it is desirable to be able to compute spinning partial waves efficiently and understand their mathematical properties. In this work we provide a systematic computational scheme valid in arbitrary spacetime dimension. Our starting point is the observation that partial waves relevant for $2\to2$ scattering of spinning particles coincide with certain matrix elements of the rotation group $SO(d-1)$. As we will show, this correspondence holds in any $d$ and for particles of any spin. The usefulness of such a relation between waves and matrix elements lies in the fact that the latter have been extensively studied in harmonic analysis. Also known as spherical functions, they have been recognised as of central importance in representation theory since the early works \cite{Gelfand-spherical,Godement1952TheoryOS,HarishChandra,Berezin-Karpelevic}. (In the context of conformal field theories, the role of spherical functions in conformal block decompositions was appreciated starting with \cite{Schomerus:2016epl,Schomerus:2017eny,Isachenkov:2018pef}.) In some selected cases, standard results about spherical functions give explicit expressions for partial waves. However, to the best of our knowledge, there seems to be no efficient way to compute these functions in generality required for applications. In the present work, we will provide such a method. The implementation in {\it Mathematica} is made freely available \href{https://gitlab.com/russofrancesco1995/partial\_waves}{ gitlab.com/russofrancesco1995/partial\_waves}.

\paragraph{Summary of results}

\smallskip

We will show that for scattering of spinning particles, the appropriate generalisation of \eqref{scalar-pw-decomposition} reads
\begin{equation}
    S = 1 + i \sum_{\pi,\rho,\sigma} a^\pi_{\rho,\sigma}(s) F^{\rho\mu}{}_{\sigma\nu}(\theta)\ w_{\rho\mu}\otimes w^{\sigma\nu}\ .
\end{equation}
As shall be explained in detail below, here $\pi$ denotes the spin of the exchanged state, $\rho$ and $\sigma$ run over spins of external two-particle states, $w^{\rho\mu}$ and $w_{\sigma\nu}$ run over bases of two-particle spin states\footnote{Analogues of polarisation vectors.} and $a^\pi_{\rho,\sigma}(s)$ are dynamical functions. The work is dedicated to the computation of partial waves $F^{\rho\mu}{}_{\sigma\nu}(\theta)$.
\smallskip

Our procedure consists of two steps. We start with partial waves for scalar scattering\footnote{Actually, from a slightly more general set of {\it seed functions}, as specified below.}, the familiar Gegenbauer polynomials. From these, spinning partial waves are generated by applications of weight-shifting operators. These operators come in two classes, depending on whether they alter quantum numbers of external or internal particles. The method is similar in spirit to \cite{Costa:2011dw,Karateev:2017jgd}, however with an important difference that all operators act directly on invariant variables (the scattering angle and {\it invariant spin variables}). What allows for computation of these reduced operators is the so-called Harish-Chandra's radial component map, \cite{HarishChandra}. Let us also note that our weight-shifting does not require knowledge of any Clebsch-Gordan coefficients. The number of applications of shift operators required to produce a spinning wave from a scalar one scales with spins of external particles and is quite small in practical applications.
\smallskip

Let us now provide more details. The most general setup that we will consider is a $2\to2$ scattering process of symmetric traceless (STT) particles in the $d$-dimensional Minkowski space $\mathbb{R}^{1,d-1}$. The particles may be massive or massless and can have equal or different spins. Since individual spins of particles are STTs, spins of two-particle states will transform in representations of their little group $SO(d-2)$ with at most two-row Young diagrams. We shall pair up particles $(12)$ and $(34)$ and label corresponding two-particle spins as $\rho=(l,\ell)$ and $\sigma=(l',\ell')$, respectively.\footnote{For simplicity, we do not consider additional multiplicity labels in this introduction.} The most general type of exchanged particles have spins with three-row Young diagrams, denoted $\pi=(J,q,s)$. We shall show that the relevant partial waves can be written as functions carrying these quantum numbers
\begin{equation}\label{functions-intro}
    f^{J,q,s}_{l,\ell,l',\ell'}(\theta,x,y)\ .
\end{equation}
It is shown that labels $(l,\ell)$ and $(l',\ell')$ are nothing else but three-point tensor structures for two external and the intermediate particle. We have traded spin indices of functions $f$ for a dependence on two spin invariants $x$ and $y$. Four-point tensor structures are written as polynomials in $x$ and $y$ of bounded degrees. With these notations in place, we can state our main results. The function \eqref{functions-intro} is an eigenfunction of the (Casimir) differential operator $\Delta^{(d)}_{l,\ell,l',\ell'}(\theta,x,y)$ given in \eqref{MST-MST-Laplacian} with the eigenvalue $C_2(J,q,s)$ given in \eqref{Casimir-Jqs}. We think of \eqref{MST-MST-Laplacian} as a family of operators parametrised by the quantum numbers. One can 'move' between different members of the family using operators $q_{l,\ell}(\theta,x,y)$ and $\bar q_{l',\ell'}(\theta,x,y)$, \eqref{WS-MST-MST-1}-\eqref{WS-MST-MST-2}, thanks to 'exchange relations' \eqref{exchange-relations}. These relations allow to shift quantum numbers $l,l'$ by one unit. A closer analysis shows that $\Delta^{(d)}_{l,\ell,l',\ell'}$ themselves can be used to shift the internal quantum numbers $(J,q,s)$. Thus, all functions \eqref{functions-intro} are obtained from the set of 'ground states', to be referred as {\it seed functions}, with very special quantum numbers,
\begin{equation}\label{ground-states}
    f^{J,\ell,0}_{\ell,0,\ell,\ell}(\theta,x,y) = \sin^\ell\theta\ C_{J-\ell}^{\left(\frac{d-3+2\ell}{2}\right)}(\cos\theta)\ .
\end{equation}
Since it is based on solving the differential equations, the weight-shifting method produces partial waves which are not normalised. For this reason, we supplement it by a normalisation procedure that is performed at the very end. The procedure is based on the simple fact that, in order to normalise any matrix element, it suffices to normalise its leading coefficient in the $\theta$-expansion. In turn, this is achieved by finite iterations of the Gelfand-Tsetlin (GT) formulas, \cite{Gelfand-Tsetlin}. After functions \eqref{functions-intro} are obtained and normalised, one obtains the partial waves $F^{\rho\mu}{}_{\sigma\nu}(\theta)$.
\smallskip

All these steps are automatically implemented in a {\it Mathematica} notebook. This means that in order to obtain a partial wave it is sufficient to plug in its quantum numbers, and the program will give back the function in an appropriate basis, the Gelfand-Tsetlin one. In Appendix \ref{Practical implementation of the algorithm}, we provide instructions for how to use the notebook. Moreover, in Section \ref{From partial waves to matrix elements} we furnish a map to go from the GT basis to the polarisation basis commonly used in the physics literature. With this map, we were able to check on the example of photon scattering in five dimensions that we get the same set of partial waves as the ones computed by other methods, \cite{correspondence}.

\smallskip
Coming back to spherical functions, it is shown that \eqref{functions-intro} are matrix elements of the $SO(d-1)$ representation $\pi$ between vectors that transform under the subgroup $SO(d-2)$ according to $\rho$ and $\sigma$. As mentioned, we are not aware of any general formula for these functions, but in special cases where results are available, our expressions agree with the literature, \cite{Vilenkin:1993:RLG2}.
\smallskip

The algebra of operators $\{\Delta,q,\bar q\}$ with appropriate potential additional members $\{p,\bar p,\dots\}$ can be constructed in a much wider scope than used in this work, with little increase in complexity of computations. Its rank-two version (rank coincides with the number of spacetime invariants, in the case at hand the single scattering angle) was constructed for purposes of conformal field theory in \cite{Buric:2022ucg}. Whereas \cite{Buric:2022ucg} shows how to increase the rank, the present work demonstrates that one can also increase the 'spin rank', i.e. the number of spin variables (cases analysed in \cite{Buric:2022ucg} had at most one spin invariant). It is expected that many physical problems are to be found among the examples covered by this more general theory, some of which will be discussed in the concluding section. Let us remark for the moment that it can be shown that a product of \eqref{functions-intro} and its analytic continuation is a defect-channel conformal block for a two-point function in the presence of a codimension-two defect. Due to the complexity of the external representations involved, these blocks go beyond the results of \cite{Lauria:2018klo,Buric:2022ucg}.
\medskip

The paper is organised as follows. In Section \ref{From partial waves to matrix elements} we introduce the setup and show how partial waves for $2\to2$ are given a group-theoretic interpretation as matrix elements. Permutation and parity symmetries are also discussed. Section \ref{Harmonic analysis} is devoted to the computation of the matrix elements. In it, we introduce the relevant tools from harmonic analysis and proceed to construct the Casimir and weight-shifting operators. The section ends with checks of our results against group theory literature. For presentation purposes, Section \ref{Harmonic analysis} focuses on the case where $\rho$ and $\sigma$ are STTs. The more general case of two-row mixed symmetry tensors is treated in Appendix \ref{External-mixed-symmetry-tensors}. The latter contains our main new results. At various points, the discussion requires the spacetime dimension to be sufficiently large, $d\geq8$, and has to be slightly adjusted for smaller dimensions. The case $d=5$ is treated in Section \ref{Exceptions in five dimensions}.\footnote{The analysis in $d=6,7$ is similar and we can provide explicit formulas upon request.} At the end of this section, we work out an example - the scattering of photons in five dimensions, and compare the resulting partial waves to known ones. The concluding Section \ref{Summary and outlook} discusses future directions and applications. Appendices \ref{Some group theory background} and \ref{Radial component map} contain background on representation theory, while \ref{External-mixed-symmetry-tensors} and \ref{From polynomials to the Gelfand-Tsetlin basis} supplement the main text by more involved calculations. Appendix \ref{Practical implementation of the algorithm} gives instructions for how to use the accompanying {\it Mathematica} code.

\section{From partial waves to matrix elements}
\label{From partial waves to matrix elements}

We begin this section by recalling some facts from Poincar\'e representation theory and introducing the relevant notation. The second and third subsection are devoted to the group-theoretic interpretation of three-point tensor structures and partial waves, respectively. In the fourth subsection, some of the general concepts are illustrated by determining the representations of two-particle states and possible exchanged particles that can appear in scattering of gravitons and photons.

\subsection{Single-particle states}
\label{Single particle states}

In this subsection, we introduce our conventions for description of massive and massless particle states. Elements of the Poincar\'e group in $d$ spacetime dimensions (we are working in mostly-positive signature)
\begin{equation}
    P = \text{ISO}(1,d-1) \cong SO(1,d-1)\ltimes \mathbb{R}^d\,,
\end{equation}
will be denoted as $(L,q)$, where $L$ is a Lorentz transformation and $q = q_\mu$ is the translation vector, $$(L,q)\cdot p = Lp + q\ .$$ We shall often refer to $p$ and $q$ as momenta (not to be confused with vectors in irreducible representations of the Poincar\'e group). The orbit of $p$ under the Lorentz group will be denoted by $\mathcal{O}_p$, and the little group of $p$ by $G_p\subset SO(1,d-1)$. Little groups of different vectors in the same orbit are related by conjugation. Physically relevant orbits, that have positive energy are either massive
\begin{equation}
    \mathcal{O}_p = \{p^\mu p_\mu = -m^2,\quad p_0>0\}, \quad \bar p=(m,0,\dots,0)\,,
\end{equation}
or massless
\begin{equation}\label{massless-orbit+standard-representative}
    \mathcal{O}_p = \{p^\mu p_\mu = 0,\quad p_0>0\}, \quad \bar p=(k,k,0,\dots,0)\ .
\end{equation}
Next to each orbit, we have written a vector belonging to it, that we shall refer to as the standard representative. The corresponding little groups are
\begin{equation}\label{little-groups}
    G_p = SO(d-1), \qquad G_p = \text{ISO}(d-2) \cong SO(d-2) \ltimes \mathbb{R}^{d-2}\ .
\end{equation}
Finally, for the description of representations, we need the notion of standard boosts. This is a family $\{\Lambda_p\}$ of Lorentz transformations labelled by momenta such that
\begin{equation}
    \Lambda_p \bar p = p\ .
\end{equation}
Standard boosts are not unique. In this work, we make the following choice for massless momenta $p$ (corresponding to the helicity basis)
\begin{equation}\label{standard-boosts-massless}
    \Lambda_p = \begin{pmatrix}
    1 & 0 & 0\\
    0 & q_2 & q_a\\
    0 & -q_a & \delta_{ab} - \frac{p_a p_b}{1+p_2}\end{pmatrix} 
    L_{01}^{\frac{k}{p^0}}, \qquad q_\alpha = \frac{p_\alpha}{p^\beta p_\beta}\ .
\end{equation}
We have written the Lorentz matrix in \eqref{standard-boosts-massless} in the block-diagonal form in the obvious way. Indices $\alpha$ and $a$ run over $\alpha=2,\dots,d-1$ and $a=3,\dots,d-1$. We refrain from writing $\Lambda_p$ in the massive case explicitly, as we will not need it in the following.
\medskip

With these notations in place, we may describe the irreducible representations of the Poincar\'e group. Consider any momentum vector of mass $m\geq0$ and let $\rho$ be an irreducible representation of the orthogonal part of its little group, \eqref{little-groups}. The Poincar\'e representation $\Pi_{m,\rho}$ is spanned by vectors $|p,w\rangle$, where $p^\mu p_\mu = -m^2$ and $w$ runs over a basis for the carrier space of $\rho$.\footnote{We use terms 'carrier space' and 'representation space' interchangeably.} The action of Poincar\'e transformations reads
\begin{equation}\label{Poincare-rerepresentaion}
    (L,q)|p,w\rangle = e^{i q_\mu (L p)^\mu} |L p, W_{Lp}(L) w\rangle, \qquad W_p(L) = \Lambda_p^{-1} L \Lambda_{L^{-1}p}\ .
\end{equation}
The Lorentz transformation $W_p(L)$ is a called a Wigner rotation. One can easily verify that \eqref{Poincare-rerepresentaion} indeed defines a representation of the Poincar\'e group.
\smallskip

The states of particles with spin are often described with the help of polarisation vectors $\epsilon^\mu$ (e.g. in \cite{Chowdhury:2019kaq,Caron-Huot:2022jli}). Thus, it is useful to provide a dictionary between our labels $w$ and polarisations. To this end, consider a massless spin-1 particle, i.e. the photon. Let $A^\mu(x) = \epsilon^\mu e^{-i p\cdot x}$ be a plane wave. We have
\begin{equation}
    (L,q)\cdot A = L\ \epsilon\ e^{-ip\cdot (L^{-1}(x-q))} = L\ \epsilon\ e^{-iLp\cdot (x-q)} = e^{iq\cdot Lp}\ L\ \epsilon\ e^{-iLp\cdot x}\ .
\end{equation}
Thus, we see that $(p,\epsilon)$ are transformed as $(Lp,L\epsilon)$. In order to achieve the correct transformation property for $w$ given in \eqref{Poincare-rerepresentaion}, one has $\epsilon = \Lambda_p (0,0,w)$.\footnote{Recall that $w\in\mathbb{R}^{d-2}$ belongs to the vector representation of $SO(d-2)$. We identify $\mathbb{R}^{d-2}$ with the last $d-2$ directions of $\mathbb{R}^{1,d-1}$. This corresponds to the choice of massless standard representative \eqref{massless-orbit+standard-representative}.} For particles of higher spin, the last relation still provides the dictionary between descriptions of states in terms of $w$-s and polarisation vectors, by means of tensor products. Indeed, in the formalism of polarisations, higher-spin particle states depend on higher-degree polynomials in $\epsilon$, which corresponds to writing vectors in higher-spin representations of $SO(d-2)$ as tensors that carry vector indices. The discussion for massive particles is entirely analogous.

\paragraph{Example} The massless little group in $d=5$ is $SO(3)\ltimes\mathbb{R}^3$. Thus, states of a photon can be labelled as $|p,w^\alpha\rangle$, where, as above, $\alpha=2,3,4$ is the vector index for $SO(3)$. Let $p=\bar p$ be the massless standard representative \eqref{massless-orbit+standard-representative}. Then the polarisation vector reads
\begin{equation*}
    \epsilon(\bar p,w) = (0,0,w^2,w^3,w^4)\ .
\end{equation*}
For a graviton, the states can be labelled as $|p,w^{\{\alpha} w^{\beta\}}\rangle$, where $\{,\}$ denotes the traceless symmetric part. In terms of polarisations, the states are quadratic polynomials in $\epsilon$. At $p=\bar p$, they are characterised by a symmetric tensor $a$
\begin{equation*}
    a_{\mu\nu} \epsilon^\mu \epsilon^\nu = a_{\alpha\beta} \epsilon^\alpha \epsilon^\beta\ .
\end{equation*}
The tracelessness condition on $a_{\alpha\beta}$ is manifested by the requirement that graviton polarisations be light-like, $\epsilon\cdot\epsilon = 0$. To compare $\epsilon$-s and $w$-s away from $p=\bar p$, one uses the standard boosts $\Lambda_p$.

\subsection{Three-point structures}
\label{Tree-point structures}

We consider a quantum field theory in $d$ spacetime dimensions, characterised by the $S$-matrix
\begin{equation}
    S = 1 + i T\ .
\end{equation}
We will focus on $2\to2$ scattering processes of spinning particles, that may be either massive or massless. The spin of any particle will be denoted by $\pi$ (we use the word spin, but $\pi$ is an arbitrary irreducible representation of the little group of an on-shell momentum). Single-particle states will be denoted by $|p,w\rangle$, where $p$ is an on-shell momentum and $w$ a vector in the representation $\pi$ of $G_p$. A two-particle state is a vector in the tensor product of two Poincar\'e irreducible representations and can be expanded in irreducible components
\begin{equation}\label{Poincare-Clebsch-Gordans}
    |p_1,w_1,p_2,w_2\rangle = \sum_{\pi,i,w} \delta(p-p_1-p_2) C^i_{12,\pi}(p,w) |p,\pi,w,i\rangle\,,
\end{equation}
where the coefficients are, by definition, the Clebsch-Gordans of the Poincar\'e group. To emphasise that the sum is over different spins $\pi$, we have included them among the labels on the right. If the decomposition is not multiplicity-free, one needs the label $i$. Regardless of whether the original particles are massive or massless, {\it most}\footnote{In this work, we consider only massive representations from the sum, i.e. massive exchanges in the scattering.} of the sum on the right hand side of \eqref{Poincare-Clebsch-Gordans} ranges over massive states. Motivated by their position space wavefunctions, we will call basis states on the left and on the right of \eqref{Poincare-Clebsch-Gordans} plane waves and spherical waves, respectively. The two-particle Hilbert space will be denoted by
\begin{equation}\label{p1-p2-into-p-decomposition}
    \mathcal{H} = \Pi_{m_1,\pi_1} \otimes \Pi_{m_2,\pi_2} \cong \int d\mu(m) \sum_{\pi,i} \Pi_{m,\pi}^{(i)}\ . 
\end{equation}
For the moment, the right-hand side is just a definition. However, let us elaborate on the on meaning of multiplicity indices $i$. Consider a three-point function, by which we mean a kinematical object of the form
\begin{equation}\label{3-pt-function}
    \langle\langle p_1, w_1, p_2, w_2| p,w\rangle\rangle = f(p_1+p_2,w_1,w_2,w)\ .
\end{equation}
Double brackets designate that the have stripped of the momentum-conserving delta function $\delta(p_1+p_2-p)$. The multiplicity of $\Pi_{m,\pi}$ inside $\Pi_{m_1,\pi_1}\otimes\Pi_{m_2,\pi_2}$ is, by definition, equal to the number of independent three-point structures. In turn, this is counted as the number of $SO(d-2)$ invariants in $\pi_1\otimes\pi_2\otimes\pi^\ast$. Indeed, let $G_{p_1,p_2}=\text{Stab}(p_1,p_2)$ be the $SO(d-2)$ subgroup of the Poincar\'e group that fixes momenta $p_1$ and $p_2$. Without loss of generality, assume that $p_{1,2}$ lie in the $01$-plane and that $p_1+p_2 = \bar p$ is the massive standard representative.\footnote{So $G_{p_1,p_2}$ is of the rotation group the subspace spanned by vectors $e_2,\dots,e_d$.} In the following, we shall refer to such a configuration as a \emph{frame}. For any $R\in G_{p_1,p_2}$ we have (see \eqref{Poincare-rerepresentaion} and \eqref{standard-boosts-frame})
\begin{equation}\label{two-particle-spins}
    R|p_1,w_1,p_2,w_2\rangle = |p_1,R w_1, p_2, R w_2\rangle\ .
\end{equation}
That is, under this group, elements labelled by spin vectors $w_1$ and $w_2$ transform as elements of the tensor product $\pi_1\otimes \pi_2$. Both for massless and massive external particles, $\pi_{1,2}$ are indeed representations of $SO(d-2)$. Furthermore, the action of $SO(d-2)$ on $w$, the spin of the third particle, is clear. The number of three-point structures is the number of $SO(d-2)$ intertwiners
\begin{equation}
    \pi_1 \otimes \pi_2 \to \pi\,,
\end{equation}
or equivalently the number of $SO(d-2)$ invariants inside $\pi_1\otimes\pi_2\otimes\pi^\ast$. The analysis of these invariants is somewhat simpler when $\pi_1\otimes\pi_2$ is multiplicity-free\footnote{The reader should be mindful of two types of multiplicities under consideration, the one for representations of $SO(d-2)$ referred to here, and the one for Poincar\'e representations discussed earlier.} (which is the case for massless particles), and we assume this from this point on. We will explain at the end how to modify the discussion to account for multiplicities, thus accounting for massive external particles as well. Since $\bar p$ is massive, $\pi$ is a representation of $SO(d-1)$. Upon restriction to $SO(d-2)$, one gets a list of irreducible representations $\pi^{(1)},\dots,\pi^{(n)}$, each of which appears in $\pi$ with multiplicity one. On the other hand, on can list irreducible components of $\pi_1\otimes\pi_2$, denoted $\pi_{12}^{(1)},\dots,\pi_{12}^{(m)}$. The number of three-point invariants can now be counted as those representations of $SO(d-2)$ that are contained in both lists. Obviously, this number depends both on external and exchanged particles. When considering partial waves, we will adopt a slightly different labelling scheme which only refers to external particles as detailed below.
\smallskip

Defining summands $\Pi^{(i)}_{m,\pi}$ on the right hand side of \eqref{p1-p2-into-p-decomposition} is equivalent to specifying three-point tensor structures. Consider a single irreducible component $\pi_{12}^{(i)}$ of $\pi_1\otimes\pi_2$ and let $P^{(i)}$ be the projector to this representation. We define in the frame
\begin{equation}\label{frame-3pt}
    \langle\langle p_1, w_1, p_2, w_2| \bar p, w,i\rangle\rangle = \langle w_1\otimes w_2 | P^{(i)} | w\rangle\,,
\end{equation}
and extend the definition by covariance to a unique three-point function.

\paragraph{Example} Consider the scattering of photons in $d=5$ dimensions. Then $\pi_1 = \pi_2 = (1)$ of the massless little group $SO(3)$. From the decomposition
\begin{equation}\label{photon-5d-two-particle-rep}
    \pi_1 \otimes \pi_2 = (1) \otimes (1) = (0) \oplus (1) \oplus (2)\,,
\end{equation}
one gets the list of two-particle spins $\{\pi_{12}^{(m)}\} = \{(0),(1),(2)\}$. Spins of intermediate particles carry two quantum numbers $(J,q)$ of $SO(4)$, which are related to two $SU(2)$-spins $(j_1,j_2)$ by $J = j_1+j_2$, $q=j_1-j_2$. By the branching rules from $SO(4)$ to $SO(3)$, see Appendix \ref{Some group theory background}, one of the $\pi_{12}^{(m)}$ from above appears in $(J,q)$ only for
\begin{equation}
    (J), \quad  (J,\pm1), \quad (J,\pm2)\ .
\end{equation}
These are the possible exchanges in the scattering process.

\subsection{Definition of partial waves}

The interacting part $T$ of the scattering matrix can be expanded in partial waves using the fact that the scattering operator $S$ commutes with generators of the Poincar\'e group. Indeed, we have
\begin{align}
    & \langle p_3,w_3,p_4,w_4|\ T\ |p_1,w_1,p_2,w_2\rangle\\
    & =\langle p_3,w_3,p_4,w_4|p',\pi',w',j\rangle\langle p',\pi',w',j|\ T\ |p,\pi,w,i\rangle\langle p,\pi,w,i|p_1,w_1,p_2,w_2\rangle\nonumber\\
    & = \delta(p'-p_3-p_4)C^j_{34,\pi'}(p',w') \langle p',\pi',w',j|\ T\ |p,\pi,w,i\rangle \delta(p-p_1-p_2)\overline{C^i_{12,\pi}(p,w)}\ , \label{T-first}
\end{align}
where the summation over internal states is understood. Since $T$ commutes with Poincare transformations, it does not change the quantum numbers
\begin{equation}
    \langle p',\pi',w',j|\ T\ |p,\pi,w,i\rangle = f^{ij}_{\pi,w}(p) \delta(p'-p)\delta_{\pi'\pi}\delta_{w'w}\ .
\end{equation}
We may therefore focus on the case $p'=p$, $\pi' = \pi$ and $w'=w$. Let $\gamma\in G_p$ be an element of the little group and $w_{1,2}$ two elements in the representation space of $\pi$ such that $w_2 = \gamma w_1$. We have
\begin{equation*}
    \langle p,\pi,w_2,j|\ T\ |p,\pi,w_2,i\rangle = \langle p,\pi,w_1,j|\ \gamma^{-1} T \gamma\ |p,\pi,w_1,i\rangle = \langle p,\pi,w_1,j|\ T\ |p,\pi,w_1,i\rangle\ .
\end{equation*}
Therefore, the coefficients $f^{ij}_{\pi,w}(p)$ do not depend on $w$. We get
\begin{align*}
    &\frac{\langle p_3,w_3,p_4,w_4|\ T\ |p_1,w_1,p_2,w_2\rangle}{\delta(p_1+p_2-p_3-p_4)}\\
    & = \sum_{\pi,w,i,j} f^{ij}_\pi(p) C^j_{34,\pi}(p,w) \overline{C^i_{12,\pi}(p,w)} = \sum_{\pi,w,i,j} f^{ij}_\pi(s) C^j_{34,\pi}(\bar p,w)  \overline{C^i_{12,\pi}(\bar p,w)}\,,
\end{align*}
where in the last step we specified to the frame $p=\bar p=(E,\vec{0})$. Partial waves are defined by performing the sum over $w$ with fixed $\pi$, $i$ and $j$
\begin{equation}\label{partial-waves-shadow-integral}
    g^{ij}_\pi = \sum_w  C^j_{34,\pi}(\bar p,w)  \overline{C^i_{12,\pi}(\bar p,w)} = \sum_w \langle\langle p_3,w_3,p_4,w_4|\bar p,\pi,w,j\rangle\rangle\langle\langle \bar p,\pi,w,i|p_1,w_1,p_2,w_2\rangle\rangle\ .
\end{equation}
We wish to regard $g^{ij}_\pi$ as functions of a single variable $\theta$, the scattering angle between incoming and outgoing particles. To this end, put $\hat p_{1i} = e_1$ and $\hat p_{3i} = e_1\cos\theta + e_2\sin\theta$. Here $e_1, e_2$ are two of the unit vectors from an orthonormal basis of $\mathbb{R}^{d-1}$ and a hat over a vector denotes unit normalisation. Then
\begin{equation}
    |p_3,w_3,p_4,w_4\rangle = e^{-\theta L_{12}} |p_1,\Lambda_{p_1}^{-1} e^{\theta L_{12}} \Lambda_{p_3} w_3,p_2,\Lambda_{p_2}^{-1} e^{\theta L_{12}} \Lambda_{p_4} w_4\rangle\,,
\end{equation}
where $\Lambda_p$ denotes the standard boost. We denote the spin components on the right by $w'_{3,4}$. Let the massive standard representative with respect to which standard boosts are defined be $p_1$. We then have (for bosonic fields)
\begin{equation}\label{standard-boosts-frame}
    \Lambda_{p_1} = 1, \quad \Lambda_{p_2} = e^{\pi L_{12}}, \quad \Lambda_{p_3} = e^{-\theta L_{12}}, \quad \Lambda_{p_4} = e^{(\pi-\theta)L_{12}}\,,
\end{equation}
and therefore $w_{3,4}'=w_{3,4}$. With these choices, the partial waves become
\begin{align*}
    g^{ij}_\pi(\theta)(w_1,\dots,w_4) & = \sum_w \langle\langle p_1,w'_3,p_2,w'_4|e^{\theta L_{12}}|\bar p,\pi,w,j\rangle\rangle\langle\langle \bar p,\pi,w,i|p_1,w_1,p_2,w_2\rangle\rangle\\
    & = \pi(e^{\theta L_{12}})^a_{\ b} \langle\langle p_1,w_3,p_2,w_4|\bar p,w^b,j\rangle\rangle\langle\langle \bar p,w_a,i|p_1,w_1,p_2,w_2\rangle\rangle\,,
\end{align*}
where $a$ is an index transforming in $\pi$.\footnote{Indices introduced in the remainder of this subsection have no relation to indices of Section \ref{Single particle states}.} Recall the transformation properties under $SO(d-2)$ of particle spins $w_{1,2}$ in the frame, \eqref{two-particle-spins}. We write the $SO(d-2)$-Clebsch-Gordan decomposition as
\begin{equation}
    |w_1\rangle \otimes |w_2\rangle = \sum_{\rho,\mu,k} \mathcal{C}^{(k)}_{12\rho} w_{(k)}^{\rho\mu}\,,
\end{equation}
where $\rho$ ranges over the representations that appear in the tensor product, $\mu$ is an index transforming in $\rho$ and $k$ accounts for multiplicities. We can label two-particle states as $|p_1,p_2,w_{(k)}^{\rho\mu}\rangle$. Using orthogonality of $SO(d-2)$ matrix elements now we obtain
\begin{align}\label{waves=matrix-elements}
     g^{ij}_\pi(\theta) (w_{(k)}^{\rho\mu}, w^{(l)}_{\sigma\nu}) &= \pi(e^{\theta L_{12}})^a_{\ b}\langle\langle p_1,p_2,w^{(l)}_{\sigma\nu}|\bar p,w^b,j\rangle\rangle\langle\langle \bar p,w_a,i|p_1,p_2,w_{(k)}^{\rho\mu}\rangle\rangle =\nonumber \\ 
     &=c^{ij}_{kl}\ \pi(e^{\theta L_{12}})^{\rho\mu}{}_{\sigma\nu}\ .
\end{align}
Therefore, up to coefficients $c^{ij}_{kl}$, partial waves coincide with the matrix elements $\pi^{\rho\mu}{}_{\sigma\nu}$. Below we show that these matrix elements are related to so-called spherical functions of the pair $(SO(d-1),SO(d-2))$.

\paragraph{Example} Let us consider the case of massive scalars. Then, there are no multiplicities and
\begin{equation}\label{zonal-section2}
    g_\pi(\theta) = \pi(e^{\theta L_{12}})^0_{\ 0} = C^{(\frac{d-3}{2})}_J(\cos\theta)\ .
\end{equation}
In the last step, we used the well-known expression for the {\it zonal spherical function}, \cite{Vilenkin:1993:RLG2}.
\smallskip

The normalisation coefficients $c^{ij}_{kl}$ are determined by the normalisation of three-point functions as we shall now explain. Under the simplifying assumption about multiplicities that we are making, indices $k$ and $l$ are absent. The partial waves are now re-written as
\begin{align}\label{matrix-elements-from-pw}
    g^{ij}_\pi(\theta)(w^{\rho\mu},w_{\sigma\nu}) & = \pi(e^{\theta L_{12}})^a_{\ b} \langle\langle p_1,p_2,w_{\sigma\nu}|\bar p,w^b,j\rangle\rangle \langle\langle \bar p,w_a,i|p_1,p_2,w^{\rho\mu}\rangle\rangle\\
    & = \pi(e^{\theta L_{12}})^a_{\ b} \langle w_{\sigma\nu}|P^{(j)}|w^b\rangle \langle w_a|P^{(i)}|w^{\rho\mu}\rangle= c(\pi,\rho,\sigma)\delta_{i\rho}\delta_{j\sigma}\ \pi(e^{\theta L_{12}})^{\rho\mu}{}_{\sigma\nu}\ .\nonumber
\end{align}
In the last step, we have used the fact that, in the frame, projectors $P^{(i)}$ project precisely to irreducible components $\rho$ in the two-particle spin space. With our choice of three-point structures specified in \eqref{frame-3pt}, we have $c(\pi,\rho,\sigma) = 1$.

\paragraph{Remark} In $d=3$, the above derivation goes through, although most of the steps become trivial. Partial waves are complex exponentials. We will not discuss this case in the following.

\subsection{Irreducible content of two-particle states}
\label{subsec:Irreducible content of two-particle states}

In previous subsections, we gave a description of three-point structures, which allow to decompose external two-particle states over internal single-particle states. The stated rules are fairly simple to use in practice to determine possible exchanged particles in any $2\to2$ scattering process. In the present subsection, we illustrate the rules on a few relevant examples.

In the first step, the tensor product of spins of external particles is decomposed into $SO(d-2)$-irreducible components to obtain the list $\{\pi_{12}^{(m)}\}$. For scalars, photons and gravitons in dimensions $d\geq6$, the decomposition reads
\begin{align*}
  & (0)\otimes(0) = (0), \qquad (1)\otimes(1) = (2)\oplus(0)\oplus(1,1),\\  &(2)\otimes(2) = (4)\oplus(2)\oplus(0)\oplus(3,1)\oplus(2,2)\oplus(1,1)\ .
\end{align*}
When $d\leq5$, the two-particle little group $SO(d-2)$ does not admit mixed symmetry tensors and these decompositions are to be modified appropriately.

After the list $\{\pi_{12}^{(m)}\}$ is obtained, one inspects for each $\pi_{12}^{(m)}$ which representations of $SO(d-1)$ contain it upon restriction to $SO(d-2)$. This is determined by the well-known $SO(d-1)\downarrow SO(d-2)$ branching rules. The rules state that a representation of $SO(d-2)$ appears in the restriction of a representation of $SO(d-1)$ if and only if a their quantum numbers satisfy a set of {\it betweenness} conditions. Concretely, if $\pi_{12}^{(m)} = (l,\ell)$ is a mixed symmetry tensor with two labels, the intermediate representation has at most three non-zero labels $(J,q,s)$, subject to
\begin{equation}
    J\geq l \geq q\geq \ell\geq s\ .
\end{equation}
Applied to scalars, photons and gravitons, these rules lead to the following sets of exchanges
\begin{align*}
    & \text{scalars} && (J)\,,\\
    & \text{photons}:  && (J), \qquad (J,q\leq2), \qquad (J,1,1)\,,\\
    & \text{gravitons}: && (J),\qquad (J,q\leq4),\qquad (J,q\leq3,1),\qquad (J,2,2)\ .
\end{align*}
As above, we stated results in generic dimension $d\geq8$. They are subject to appropriate modifications in low dimensions.\footnote{See Section \ref{Matrix elements as spherical functions} for details.} The first line is the familiar statement that scalar particles exchange symmetric traceless tensors of arbitrary spin $J$. After possible intermediate particles are determined, one proceeds to count the three-point tensor structures, i.e. multiplicities, as explained in the last subsection - $\pi=(J,q,s)$ is restricted to $SO(d-2)$ and the result compared to the list $\{\pi_{12}^{(m)}\}$. For external scalars, the number of tensor structures is always one. For photons, possible multiplicities are one and two, corresponding to exchanges
\begin{align*}
    & \text{multiplicity two}: \quad (J\geq2), \quad (J\geq2,1)\,,\\
    & \text{multiplicity one}: \quad (1),\quad (0), \quad (1,1), \quad (J,2), \quad (J,1,1)\ .
\end{align*}
For gravitons, the multiplicities go up to four
\begin{align*}
    & \text{multiplicity four}: \quad (J\geq4,2), \quad (J\geq4,1)\,\\
    & \text{multiplicity three}: \quad (J\geq4),\quad (3,2), \quad (3,1)\,\\
    & \text{multiplicity two}: \quad (3), \quad (2),\quad (J\geq4,3), \quad (2,2), \quad (2,1), \quad (J\geq3,2,1)\,\\
    & \text{multiplicity one}: \quad (1),\quad (0), \quad (J,1), \quad (3,3), \quad (1,1)\,,\\
    & \hskip3.3cm (J,3,1), \quad (2,2,1), \quad (2,1,1), \quad (J,2,2)\ .
\end{align*}
Above results do not account for parity or the fact that particles in the scattering process are identical. These additional symmetries in general restrict the possible exchanges and reduce the number of three-point structures. We turn to them presently.

\paragraph{Remark} Another view on tensor product decompositions of Poincar\'e particles is provided by the {\it tensor product theorem}, \cite{raczka1986theory}. The theorem determines intermediate spins, counted with multiplicity, as $SO(d-2)$-irreducible components of the induced representation
\begin{equation}
    \text{Ind}_{SO(d-2)}^{SO(d-1)}(\pi_1\otimes\pi_2)\ .
\end{equation}
One may verify that this succinct description is equivalent to the rules we stated above. The equivalence between the two is a manifestation of Frobenius reciprocity, \cite{Kirillov}, which relates notions of induced and restricted representations. 

\paragraph{Summary} We summarise the procedure arrived at above, focusing on massless STT particles. Other cases are similar. Start with the set of external spins $(J_1,J_2,J_3,J_4)$. Compute the external irreducible representations (the vertices) by the $SO(d-2)$-tensor decomposition,
\begin{equation}
    (J_1) \otimes (J_2) = \bigoplus_i \rho_i \qquad (J_3) \otimes (J_4) = \bigoplus_j \sigma_j \, .
\end{equation}
For every pair of vertices $(\rho,\sigma)$, list the representations $\pi$ of $SO(d-1)$ that contain both $\rho$ and $\sigma$ upon restriction to $SO(d-2)$ (intermediate representations). For every triple $(\rho,\sigma,\pi)$, there is the corresponding partial wave $\pi^\rho{}_\sigma$. It is a vector-valued function and coincides with the set of matrix elements $\pi(e^{\theta L_{12}})^{\rho\mu}{}_{\sigma\nu}$. The latter are computed using the implementation described in Appendix \ref{Practical implementation of the algorithm}. Schematically,

\begin{center}
\begin{tikzpicture}
    \node (f) at (0,0) {$(J_1,J_2,J_3,J_4)$}; 
    \node (g) at (3,0) {$\{\rho_i\}, \, \{\sigma_j\}$};  
    \node (h) at (6,0) {$\forall \, (\rho,\sigma)$ form $\{\pi\}$ };
    \node (i) at (10,0) {compute $\pi^\rho{}_\sigma \, .$};

    \draw[->] (f) -- (g);
    \draw[->] (g) --  (h);
    \draw[->] (h) -- (i);
\end{tikzpicture}
\end{center}

\section{Harmonic analysis}
\label{Harmonic analysis}

This section is dedicated to a detailed study of matrix elements \eqref{matrix-elements-from-pw}. After illustrating the main ideas on the example of $SO(3)$ in the first subsection, we proceed to give general definitions related to matrix elements and spherical functions in Section \ref{Matrix elements as spherical functions}. In the next two subsections, we derive the reduced Laplace and weight-shifting operators acting on the spherical functions. Section \ref{solution theory STT-STT} develops the solution theory for these functions. A subset of our results is checked against the known literature in Section \ref{subsec: Matrix elements in the Gelfand-Tsetlin basis and checks}. For simplicity of presentation, the present section only contains explicit calculations with the assumption that $\rho$ and $\sigma$ are STTs. The more general case with $\rho$ and $\sigma$ two-row MSTs is treated in Appendix \ref{External-mixed-symmetry-tensors}.
\smallskip

Unless specified otherwise, we will denote the three groups relevant for the discussion by
\begin{equation}
    G = SO(d-1), \quad K = SO(d-2), \quad M = SO(d-3)\ .
\end{equation}
For the most part, the global structure of the groups will not play a role, because we are interested in differential operators, which are local objects. The dependence on global properties only enters through possible choices for representations of $G$ and $K$.
\smallskip

The Lie algebra of $G$ is denoted by $\mathfrak{g}$, and similarly for other groups. The Lie algebra $\mathfrak{g}$ is spanned by elements $\{L_{AB}\}$ with brackets
\begin{equation}
    [L_{AB},L_{CD}] = \delta_{BC} L_{AD} - \delta_{AC} L_{BD} + \delta_{BD} L_{CA} - \delta_{AD} L_{CB}, \quad A,\dots,D = 1,\dots,d-1\ .
\end{equation}
When attached to $L$, indices $A\dots$ will always run over these values. We will use indices $\mu,\nu=2,\dots,d-1$ and $i,j=3,\dots,d-1$. Thus, $\{L_{\mu\nu}\}$ are generators of $K$ and $\{L_{ij}\}$ are generators of $M$. We will re-use some types of indices to label bases of various representation spaces. In any equation, which of these two meanings is being used should be clear from the context.

\paragraph{Remark} The Riemannian Laplace-Beltrami operator $\Delta$ on $G$ with respect to the bi-invariant metric coincides with the quadratic Casimir constructed from invariant vector fields. Either left- or right-invariant vector fields may be used, as they lead to the same operator. We will use terms 'group Laplacian', 'Laplacian' and 'Casimir' interchangeably to refer to this operator.

\subsection{Illustration: Wigner-$d$ functions}
\label{Illustration Wigner d}

Before discussing matrix elements \eqref{matrix-elements-from-pw} in full generality, we illustrate their properties on the simplest example, that of the group $SO(3)$. By the previous section, these functions coincide with partial waves in $d=4$ dimensions. 

\vskip0.1cm The group $G = SO(3)$ is conveniently parametrised by Euler angles $(\phi,\theta,\psi)$
\begin{equation}\label{Euler-angles-SU(2)}
g(\phi,\theta,\psi) = e^{\phi L_{23}} e^{\theta L_{12}} e^{\psi L_{23}}\ .
\end{equation}
Irreducible matrix elements of $G$ are dense in the space of functions $L^2(G)$. Let $|j,m\rangle$ be eigenstates of $L_{23}$ in the spin-$j$ representation. In the Euler angle parametrisation, matrix elements between such states factorise into a product of two exponentials and the Wigner-$d$ function\footnote{Compare to our previous notation in \eqref{matrix-elements-from-pw}, $\pi \sim (j)$ and $\rho,\sigma\sim m,m'$, the indices $\mu,\nu$ being redundant.}
\begin{equation}\label{matrix-elements-SU(2)}
    \langle j,m| e^{\phi L_{23}} e^{\theta L_{12}} e^{\psi L_{23}}| j',m'\rangle = \delta_{jj'} e^{-i(m\phi-m'\psi)} d^j_{mm'}(\theta)\ .
\end{equation}
One possible way to determine the Wigner-$d$ function is by imposing that the function on the right hand side of \eqref{matrix-elements-SU(2)} is an eigenfunction of the group Laplacian with the eigenvalue $-j(j+1)$. The Laplacian is the operator in all group coordinates
\begin{equation}\label{Laplacian-SU(2)}
    \Delta = \partial_\theta^2 + \cot\theta\ \partial_\theta + \frac{\partial_\phi^2 - 2\cos\theta\partial_\phi\partial_\psi + \partial_\psi^2}{\sin^2\theta}\ .
\end{equation}
When acting on matrix elements \eqref{matrix-elements-SU(2)}, we can substitute $\partial_\phi\to-im$, $\partial_\psi\to im'$ to reduce $\Delta$ to a single-variable operator that acts on $d^j_{mm'}(\theta)$, namely
\begin{equation}\label{delta-mn}
    \Delta_{m,m'} = \partial_{\theta}^2 + \cot\theta\ \partial_\theta - \frac{m^2 + 2m m'\cos\theta + m'^2}{\sin^2\theta}\ .
\end{equation}
Solving the eigenvalue equation $\Delta_{m,m'}d^j_{mm'} = -j(j+1)d^j_{mm'}$, one gets the familiar expression for the $d$-function in terms of Jacobi polynomials. In the case $m=m'=0$, these reduce to Legendre polynomials, $d^j_{00}(\theta) = P_j(\cos\theta)$, as can be directly seen from the change of variables $t=\cos\theta$\footnote{We will use the notation $t=\cos\theta$ for the remainder of the text. It is not to be confused with a Mandelstam variable.} that maps \eqref{delta-mn} to the Legendre equation.
\smallskip

In higher dimensions, the above derivation meets some difficulties. Firstly, our computation of the reduced Laplacian $\Delta_{m,m'}$ passed through that of the more complicated operator $\Delta$ in all group coordinates. Secondly, for $d>4$ the group $K=SO(d-2)$ is no longer abelian and its (non-trivial) irreducible representations are not one-dimensional. Therefore, matrix elements have further dependence on states in $K$-representations. Below, both of these problems will be resolved using a tool called Harish-Chandra's radial component map, \cite{HarishChandra}. The map allows to obtain the reduction of the Laplacian to any space $K$-$K$-covariant matrix elements by a simple manipulation in the universal enveloping algebra of $\mathfrak{g}$. In the remainder of this subsection, we describe the general prescription on the above example.
\smallskip

In the first step, we introduce a new basis for $\mathfrak{g}$
\begin{equation}
    \mathfrak{g} = \text{span}\{L_{12}, L_{23}, L'_{23}\}, \qquad L_{23}' = e^{-\theta L_{12}} L_{23} e^{\theta L_{12}} = \cos\theta\ L_{23} - \sin\theta L_{13}\ .
\end{equation}
Any polynomial in the generators $L_{12}$, $L_{13}$, $L_{23}$ may also be written as a polynomial in $L_{12}$, $L_{23}$, $L'_{23}$. In writing such polynomials, we impose the ordering prescription in which $L'_{23}$ and $L_{23}$ always appear on the left and the right of $L_{12}$, respectively. Any element of $U(\mathfrak{g})$ written according to these rules is said to be radially decomposed. The radial decomposition of the quadratic Casimir reads
\begin{equation}\label{radial-decomposition-SU(2)}
    C_2 = L_{12}^2 + \cot\theta\ L_{12} + \frac{L'^2_{23} - 2\cos\theta\ L'_{23} L_{23} + L^2_{23}}{\sin^2\theta}\ .
\end{equation}
We observe that the reduced Laplacian $\Delta_{m,m'}$ is closely related to the radial decomposition of $C_2$. Namely, the former is obtained from the latter by making substitutions
\begin{equation}\label{substitutions}
    L_{12}\to\partial_\theta,\quad L'_{23}\to -im, \quad L_{23}\to im'\ .
\end{equation}
This result is a special case of Harish-Chandra's theorem. The theorem can also be suitably applied for manipulations of non-Casimir elements of $U(\mathfrak{g})$, and in particular the generators (elements of $\mathfrak{g}$) themselves. Let us radially decompose the raising and lowering operators for $L_{23}$
\begin{equation}
    L = L_{12} + i L_{13} = L_{12} + i\cot\theta\ L_{23} - \frac{i L'_{23}}{\sin\theta}, \quad \bar L = L_{12} - i L_{13} = L_{12} - i\cot\theta\ L_{23} + \frac{i L'_{23}}{\sin\theta}\,,
\end{equation}
and apply the same substitutions \eqref{substitutions}. This gives rise to operators
\begin{equation}
    q_{m,m'} = \partial_\theta - m' \cot\theta - \frac{m}{\sin\theta}, \quad \bar q_{m,m'} = \partial_\theta + m'\cot\theta + \frac{m}{\sin\theta}\ .
\end{equation}
Operators $q_{m,m'}$ and $\bar q_{m,m'}$ can be used to raise and lower the index $m'$ on $\Delta_{m,m'}$ by one, thanks to the exchange relations
\begin{equation}
    \Delta_{m,m'+1}q_{m,m'} = q_{m,m'} \Delta_{m,m'}, \quad \Delta_{m,m'-1}\bar q_{m,m'} = \bar q_{m,m'} \Delta_{m,m'}\ .
\end{equation}
There is a similar pair of operators that change the value of $m$. Therefore, through applications of these shifting operators one obtains all matrix elements from Legendre polynomials $d^j_{00}(\theta)$.\footnote{By using the expression for the $d$-function in terms of Jacobi polynomials, acting with $q$ and $\bar q$ leads to certain identities satisfied by these polynomials. Concretely, see equations (6) and (7) on page 335 of \cite{Vilenkin}.}

Harish-Chandra's theorem applies in higher dimensions and for matrix elements transforming in arbitrary representations of $K$. Therefore, it allows for a simple computation of Casimir and shifting operators. When applied in appropriate order to certain {\it ground states}, higher-dimensional analogues of Legendre polynomials, these operators generate all matrix elements. The present section carries out this construction.

\subsection{Matrix elements and spherical functions}
\label{Matrix elements as spherical functions}

Let $\pi$ be an irreducible representation of $G$ and $\rho,\sigma$ irreducible representations of $K$. Denote by $V$ the carrier space of $\pi$ and its basis by $\{e_i\}$. We will be studying particular classes of matrix elements of $\pi$, i.e. the complex-valued functions on the group
\begin{equation}
    \pi^i_{\ j}(g) = \langle e^i | \pi(g) | e_j \rangle\ .
\end{equation}
Let $W_l$ and $W_r$ be the carrier spaces of $\rho$ and $\sigma$ and denote their bases by $\{e_a\}$ and $\{e_\alpha\}$, respectively.\footnote{In general, representations $\rho$ and $\sigma$ are non-isomorphic, so we use two types of indices.} We will use the Dirac notation and write basis elements of $W_l\otimes W_r^\ast$ as $|e_a\rangle\langle e^\alpha|$. Functions we are interested in are
\begin{equation}\label{matrix-element}
    F^a_{\ \alpha}(\theta) = \langle e^a| \pi(e^{\theta L_{12}}) |e_\alpha\rangle\ .
\end{equation}
The definition only makes sense if $\pi$, when restricted to $K$, contains representations $\rho$ and $\sigma$. Then $W_l,W_r\subset V$ and vectors $e_a$, $e_\alpha$ are elements of $V$. This will be assumed in the rest of the text. It is important to notice that if $\rho$ appears in $\pi|_K$, it does so with multiplicity one. This is a general property of restrictions between consecutive orthogonal groups, encapsulated by saying that the restriction from $SO(d-1)$ to $SO(d-2)$ is multiplicity-free.
\smallskip

We now wish to establish covariance properties of $\pi^a_{\ \alpha}$ under left and right multiplication of its argument by elements of $K$. We have
\begin{equation*}
    \pi^a_{\ \alpha}(k_l g k_r) = \langle e^a|\pi(k_l)|e_i\rangle\langle e^i|\pi(g)|e_j\rangle\langle e^j|\pi(k_r)|e_\alpha\rangle\ .
\end{equation*}
By the orthogonality of matrix elements of $K$, we get non-zero contributions only from $i=b$ and $j=\beta$ (multiplicity-freeness means that this notion is well-defined, i.e. there is only one $e_b$ among $e_i$-s)
\begin{equation}\label{covariance-laws}
    \pi^a_{\ \alpha}(k_l g k_r) = \rho^a_{\ b}(k_l) \pi^b_{\ \beta}(g) \sigma^\beta_{\ \alpha}(k_r)\ .
\end{equation}
Functions satisfying \eqref{covariance-laws} are called spherical. We will call $\pi^a_{\ \alpha}$ or $f^a_{\ \alpha}$ a spherical function, but it should be understood that this is a vector-valued function, with two indices. The space of spherical functions depends on the two representations $\rho$, $\sigma$ - it will be denoted by $\Gamma_{\rho,\sigma}$
\begin{equation}
    \Gamma_{\rho,\sigma} = \{f : G\to\text{Hom}(W_r,W_l)\ |\ f(k_l g k_r) = \rho(k_l) f(g) \sigma(k_r)\}\ .
\end{equation}
We now specialise the preceding abstract discussion to cases of interest. There are in total three representations that play a role. These are external representations $\rho$ and $\sigma$ of $SO(d-2)$ and the internal representation $\pi$ of $SO(d-1)$. The three are not completely independent - upon restriction to $SO(d-2)$, $\pi$ contains both $\rho$ and $\sigma$.

\paragraph{Synopsis of $SO(n)$ representations} Before continuing, we briefly review elements of representation theory of $SO(n)$. Let us denote the rank of this group by $r$, thus $n=2r$ or $n=2r+1$. Irreducible representations of $SO(n)$ can be labelled by sequences $(l_1,\dots,l_r)$ satisfying
\begin{align}
    & SO(2r+1): \quad l_1\geq l_2\geq\dots\geq l_r\geq 0\,,\\
    & SO(2r): \ \ \ \ \ \quad l_1\geq\dots\geq l_{r-1}\geq |l_r|\ .
\end{align}
In addition, either all of $l_i$ are integers (for bosonic representations), or they are all half-integers (for fermionic representations of the double cover Spin$(n)$). In the integer case, $l_i$ are the lengths of rows in the associated Young diagram.
\smallskip

The Gelfand-Tsetlin (GT) patterns are a particular choice of basis vectors for an $SO(n)$ representation. The labelling scheme is based on the fact that restrictions from $SO(n)$ to $SO(n-1)$ are multiplicity free. Indeed, upon restriction to $SO(2r)$, the representation $(l_1,\dots,l_r)$ of $SO(2r+1)$ contains the representation $(m_1,\dots,m_r)$ if and only if the betweenness conditions hold
\begin{equation}
    l_1 \geq m_1 \geq l_2 \geq \dots \geq l_r \geq |m_r|\,,
\end{equation}
and each of the allowed representations $(m_1,\dots,m_r)$ appears with multiplicity one. A similar rule applies to restrictions from $SO(2r)$ to $SO(2r-1)$, see e.g. \cite{Dobrev:1977qv}. Proceeding inductively, one can describe a unique vector (up to normalisation) in the representation space $(l_1,\dots,l_r)$ of $SO(n)$ by its transformation properties under $SO(n-1),SO(n-2),\dots,SO(2)$. The associated sets of labels taken together form a GT pattern.

\bigskip

For the remainder of this section, we shall assume that $\rho$ and $\sigma$ are symmetric traceless tensors of spin $l$ and $l'$, respectively (see Appendices \ref{External-mixed-symmetry-tensors} and \ref{From polynomials to the Gelfand-Tsetlin basis} for the analysis when $\rho$ and $\sigma$ are two-row MSTs). This means that intermediate representations have only two non-vanishing Gelfand-Tsetlin labels, denoted $(J,q)$. Furthermore, the labels are constrained as
\begin{equation}\label{selection-rules}
    J \geq l,l' \geq q\ .
\end{equation}
The quadratic Casimir value in this representation is, \cite{Dobrev:1977qv},
\begin{equation}\label{Casimir-values}
    C_2(\pi_{J,q}) = -J(J+d-3) - q(q+d-5)\ .
\end{equation}
Now assume that $l$, $l'$, $J$ and $q$ are fixed. We ask over which sets the vectors $\{e^a\}$ and $\{e_\alpha\}$ run to produce a non-zero matrix element \eqref{matrix-element}. First of all, $e^a$ is represented by an $SO(d-1)$ Gelfand-Tsetlin pattern whose first row is fixed at $(J,q,0,\dots,0)$ and the second row at $(l,0,\dots,0)$. All the remaining rows have at most one non-zero element. We denote these by
\begin{equation}
    l\geq m_{d-3} \geq m_{d-4} \geq \dots \geq m_2\geq 0\ .
\end{equation}
Similar comments apply to the vector $e_\alpha$, which is specified by a sequence $m'_{d-3},\dots,m'_2$. Since $e^{\theta L_{12}}$ commutes with $M$, non-zero matrix elements have $m'_i = m_i$. Thus, we are left with
\begin{equation}\label{equal-mmp}
    f = \langle l;m_{d-3}\dots m_2| \pi_{J,q}(e^{\theta L_{12}})|l'; m_{d-3}\dots m_2\rangle\ .
\end{equation}
However, not all of these matrix elements are independent. In fact, they depend only on $m_{d-3}$. The proof is a version of Schur's lemma. Consider two matrix elements, \eqref{equal-mmp} and
\begin{equation}
    \tilde f = \langle l;m_{d-3}\tilde m_{d-4}\dots \tilde m_2| \pi_{J,q}(e^{\theta L_{12}})|l'; m_{d-3}\tilde m_{d-4}\dots \tilde m_2\rangle\ .
\end{equation}
The label $m_{d-3}$ specifies an irreducible representation of $SO(d-3)$. The remaining labels define two vectors in this representation, $v=|m_{d-4}\dots m_2\rangle$ and $\tilde v=|\tilde m_{d-4}\dots \tilde m_2\rangle$. Since the representation is irreducible, there is an element $\gamma\in SO(d-3)$ such that $\tilde v= \gamma v$. Therefore, we get
\begin{equation*}\label{Schur}
    \tilde f = \langle l,m_{d-3},\tilde v|\pi_{J,q}(e^{\theta L_{12}})| l',m_{d-3},\tilde v\rangle = \langle l,m_{d-3},v|\pi_{J,q}(\gamma^{-1}e^{\theta L_{12}}\gamma)| l',m_{d-3},v\rangle = f\ .
\end{equation*}
In the last step, we have used that $\gamma$ commutes with $e^{\theta L_{12}}$. In conclusion, we may label the independent matrix elements as
\begin{equation}\label{matrix-elements-final}
    F^{d-1,J,q}_{l,l',j}(\theta) = \langle l,j, v|\pi_{J,q}(e^{\theta L_{12}})| l',j, v\rangle\ .
\end{equation}
For most of our discussion, the dimension $d$ is fixed, and we will drop the corresponding label on $F$. The label $j$ runs from 0 to $\text{min}(l,l')$, i.e. over representations of $SO(d-3)$ that are contained in both $(l)$ and $(l')$ of $SO(d-2)$. Thus, the number of independent functions is
\begin{equation}\label{number-structures}
    N(l,l') = \text{min}(l,l') + 1\ .
\end{equation}
If either $l$ or $l'$ vanishes, matrix elements \eqref{matrix-elements-final} are called \emph{associated spherical functions}. If $q=0$, they are called spherical functions \emph{of class I}. Integral representations for matrix elements of class I can be found in \cite{Vilenkin}. The remainder of this section is devoted to the computation of \eqref{matrix-elements-final}. The extension to cases where $\rho,\sigma$ are two-row mixed symmetry tensors is given in Appendices \ref{External-mixed-symmetry-tensors} and \ref{From polynomials to the Gelfand-Tsetlin basis}.

\subsection{Differential equations from the radial component map}
\label{Differential equations from the radial component map}

Matrix elements $\pi^i_{\ j}$ of unitary irreducible representations are eigenfunctions of the Laplacian and higher Casimir operators. The eigenvalues are equal to values of Casimirs in the representation $\pi$, e.g.
\begin{equation}
    \Delta \pi^i_{\ j} = C_2(\pi) \pi^i_{\ j}\ .
\end{equation}
We wish to use this equation to compute spherical functions. However, as it stands, $\Delta$ is a complicated differential operator in the $\text{dim}(G)$ coordinates that parametrise the group. 
\smallskip

Because it commutes with left and right invariant vector fields, the Laplacian preserves each space of spherical functions $\Gamma_{\rho,\sigma}$. Functions in $\Gamma_{\rho,\sigma}$ may be regarded as depending on a single variable, due to the so-called Cartan decomposition of $G$. This is a higher-dimensional generalisation of the Euler angle decomposition for $SO(3)$ and states that any element $g\in G$ can be factorised as
\begin{equation}\label{Cartan}
    g = k_l\ e^{\theta L_{12}}\ k_r, \qquad k_l,k_r\in K\ .
\end{equation}
It is customary to denote the middle factor by $a=e^{\theta L_{12}}$. The one-dimensional subgroup generated by $L_{12}$ will be denoted by $A_p$. The Cartan decomposition of an element $g$ is far from unique. Let $m\in M$ and notice that $M$ consists of those elements in $K$ that commute with $e^{\theta L_{12}}$. Given one factorisation \eqref{Cartan}, we can produce another of the same form
\begin{equation}
    g = (k_l m)\ e^{\theta L_{12}}\ (m^{-1} k_r)\ .
\end{equation}
If $\pi^a_{\ \alpha}$ is a spherical function, once we know $\pi^a_{\ \alpha}(e^{\theta L_{12}})$, left and right covariance properties allow us to extend it to the whole of $G$. However, not any function on $A_p$ can be extended to a spherical function on $G$. Indeed
\begin{equation}\label{m-invariance}
    \pi(a) = \pi(m a m^{-1}) = \rho(m) \pi(a) \sigma(m^{-1})\ .
\end{equation}
Thus, $F = \pi|_{A_p}$ takes values in the subspace of vectors of $W_l\otimes W_r^\ast$ which satisfy the invariance condition \eqref{m-invariance}. We will exhibit this condition in more concrete terms in examples below.
\smallskip

We now come to the main point of the present section, namely that the Laplacian eigenvalue problem can be written as an explicit differential equation for the restriction $F$ of $\pi$ to $A_p$. As we mentioned above if $f\in\Gamma_{\rho,\sigma}$, also $\Delta f\in\Gamma_{\rho,\sigma}$. This means that $\Delta$ can be reduced to a single-variable differential operator $\Delta_{\rho,\sigma}$, which depends on choices of $\rho$ and $\sigma$. The operator is explicitly given by
\begin{equation}\label{main}
    \Delta_{\rho,\sigma} = \partial_{\theta}^2 + (d-3)\cot\theta\ \partial_\theta + \sum_{i=3}^{d-1} \frac{\rho(L^2_{2i}) + 2\cos\theta\ \rho(L_{2i})\sigma^\ast(L_{2i}) + \sigma^\ast(L_{2i}^2)}{\sin^2 \theta} + \frac12 \sigma^\ast(L^{ij} L_{ij})\ .
\end{equation}
Notice that arguments of $\rho$ and $\sigma^\ast$ in this expressions are polynomials in elements of the Lie algebra of $K$, $\mathfrak{k}$, and thus the formula \eqref{main} is well-defined. The object on the right hand side is a differential operator in $\theta$ with coefficients in $\text{End}(W_l\otimes W_r^\ast)$. E.g. if representations $\rho$ and $\sigma$ are realised by matrices, $\Delta_{\rho,\sigma}$ becomes a differential operator with a matrix potential term. The proof of \eqref{main} follows from the existence of Harish-Chandra's radial component map. The result of Harish-Chandra, a special case of which we discussed in Section \ref{Illustration Wigner d}, states that reductions of any Casimir operator to spaces of spherical functions can be performed universally in spin $\rho$ and $\sigma$ once the Casimir is {\it radially decomposed} in the universal enveloping algebra of $\mathfrak{g}$. We give some more details about the radial component map and how it leads to \eqref{main} in Appendix \ref{Radial component map}.
\smallskip

We will now assume that both $\rho$ and $\sigma$ are STTs of $SO(d-2)$. These representations are conveniently realised on the space of polynomials in variables $x^a$ of degree $\leq l$, where $a=4,...,d-1$
\begin{align}\label{differential-basis-1}
    & \rho(L_{23}) = -i( x_a\partial_a - l),\quad \rho(L_{2a}) = -\frac12 \Big((1-x^2)\partial_a + 2x_a (x^b\partial_b - l)\Big)\,,\\
    & \rho(L_{3a}) = \frac{i}{2}\Big((1+x^2)\partial_a - 2x_a(x^b\partial_b-l)\Big)  ,\quad  \rho(L_{ab}) = x_a \partial_b - x_b\partial_a\ .\label{differential-basis-2}
\end{align}
Similarly, $\sigma^\ast$ is realised in terms of operators in coordinates $x'^a$ and has spin $l'$. Invariance condition related to the stabiliser $M$ written in \eqref{m-invariance} implies that the differential operator $\Delta_{\rho,\sigma}$ is applied to functions $F(x^a,x'^a)$ that satisfy
\begin{equation}\label{constraints}
    (x_a \partial_b - x_b\partial_a + x'_a \partial'_b - x'_b\partial'_a) F(x^a,x'^a) = 0\ .
\end{equation}
The condition is satisfied precisely by functions of scalar products
\begin{equation}
    F = F(x^a x_a, x'^a x'_a,x^a x'_a) \equiv F(X,X',W)\ .
\end{equation}
While these conditions are the simplest ones to implement, \eqref{m-invariance} additionally requires
\begin{equation}\label{SO(d-4)-invariance}
   (\rho(L_{3a}) + \sigma^\ast(L_{3a})) F = 0\ .
\end{equation}
The general solution to all the constraints reads
\begin{equation}\label{form-of-functions-STT}
    F = (1-X)^l (1-X')^{l'} f(y), \qquad y = \frac{(X+1)(X'+1)-4W}{(X-1)(X'-1)}\ .
\end{equation}
The operators $\rho(L_{2i}L_{2i})$, $\sigma^\ast(L_{2i}L_{2i})$ and $\rho(L_{2i})\sigma^\ast(L'_{2i})$ all commute with the constraints \eqref{constraints}, \eqref{SO(d-4)-invariance} and therefore reduce to well-defined operators in $y$. To write the final result, put
\begin{equation}\label{Gegenbauer-operator}
    \mathcal{D}^{(d)}_y = (y^2-1)\partial_y^2 + (d-4)y\partial_y\ .
\end{equation}
Notice that this is the Gegenbauer differential operator, i.e. Gegenbauer polynomials satisfy the differential equation
\begin{equation}\label{Gegenbauer-equation}
    \mathcal{D}^{(d)}_y C^{(\frac{d-5}{2})}_n(y) = n(n+d-5)  C^{(\frac{d-5}{2})}_n(y)\ .
\end{equation}
The reduced Laplacian reads
\begin{align}\label{final}
    & \Delta^{(d)}_{l,l'} = \partial_{\theta}^2 + (d-3)\cot\theta\ \partial_\theta -\mathcal{D}^{(d)}_y +\frac{2\mathcal{D}^{(d)}_y - l(l+d-4) - l'(l'+d-4)}{\sin^2\theta} \\
    & - 2\cos\theta\ \frac{ y\mathcal{D}^{(d)}_y - (l+l'+d-5)(y^2-1)\partial_y + l l' y}{\sin^2\theta}\ .\nonumber
\end{align}
Equation \eqref{final} is one of the main results of the present section and we shall often return to it. In summary, the family of operators \eqref{final} are all possible reductions of the group Laplacian to spaces of spherical functions with symmetric traceless left-right covariance laws, $\Gamma_{(l),(l')}$. The appropriate generalisation of \eqref{final} that includes covariance laws for mixed symmetry tensors is written in \eqref{MST-MST-Laplacian}. 
\smallskip

If we go back and regard $F$ as a vector valued function of $\theta$, its number of components is equal to the number of $M$-invariants in $\rho\otimes\sigma^\ast$. Without loss of generality, assume that $l'\geq l$. Let us see how \eqref{number-structures} appears in the differential operator $\Delta^{(d)}_{l,l'}$. Clearly, this operator maps polynomials in $y$ again to polynomials and increases their degree by at most one. The term that increases the degree is the one multiplied by $\cos\theta/\sin^2\theta$. However, one checks explicitly that
\begin{equation}
     \Delta^{(d)}_{l,l'} \left(f(\theta)y^l\right) = f_1(\theta)y^l + f_2(\theta)y^{l-1}\ .
\end{equation}
Therefore, $\Delta^{(d)}_{l,l'}$ preserves the space of functions $f(\theta,y)$ which are polynomials in $y$ of degree less than or equal to $l$. By expanding in $y$, we get a function of $\theta$ with $l+1 = N(l,l')$ components.

\paragraph{Remark} Invariance conditions \eqref{m-invariance} correspond to four-point invariance of partial waves \eqref{spaces-3pt-4pt-structures} and powers of $y$ give a particular choice of four-point tensor structures.

\subsection{Weight-shifting operators}
\label{subsec:Weight-shifting operators}
To construct spherical functions with complicated internal and external representations, it is often useful to start with ones with simpler representations, and apply to them {\it weight-shifting operators}. In the following subsections, we shall explore two constructions of such shifting operators, which, depending on the representations they change, shall be referred to external and internal.

\subsubsection{External weight-shifting operators}

Before going on, let us make a few remarks about invariant vector fields. Any element $X\in\mathfrak{g}$ gives rise to left- and right-invariant vector fields on $G$, $\mathcal{L}_X$ and $\mathcal{R}_X$, the generators of right and left regular actions, respectively. Given a local coordinate system $(x^i)$ on the group and a basis $\{X^j\}$ of its Lie algebra, the vector fields are written as differential operators
\begin{equation}
    \mathcal{L}_{X^i} = C^{ik}_L(x) \partial_{x^k},\quad \mathcal{R}_{X^i} = C^{ik}_R(x) \partial_{x^k}\ .
\end{equation}
The coefficient functions may be computed from the left and right Maurer-Cartan forms $g^{-1}dg$ and $dg g^{-1}$. However, the only property of vector fields that is of interest for the present discussion is that they form two representations of the Lie algebra $\mathfrak{g}$, and commute with one another. More precisely, in accord with conventions of \cite{Buric:2022ucg}, left-invariant fields satisfy the same brackets as Lie algebra generators, while right-invariant vector fields satisfy the opposite brackets.

Vector fields act on spherical functions component-wise. Infinitesimally, the covariance conditions \eqref{covariance-laws} read
\begin{equation}
    \mathcal{R}_k f^a_{\ \alpha} = \rho^a_{\ b}(k) f^b_{\ \alpha}, \quad \mathcal{L}_k f^a_{\ \alpha} = f^a_{\ \beta} \sigma^\beta_{\ \alpha}(k), \quad k\in \mathfrak{k}\ .
\end{equation}
Unlike Casimir operators, vector fields do not commute with generators of $K$, and therefore do not preserve spaces $\Gamma_{\rho,\sigma}$. Our idea is not to apply these operators to spherical functions individually. Rather, the vector fields are first collected according to their transformation properties under $K$ and then applied "collectively". In this way, while the $K$-covariance properties of the original function are altered, they are so in a definite way and we end up with a spherical function from a different space $\Gamma_{\rho',\sigma'}$.
\smallskip

Let us now be more precise. Generators $L_{1\mu}$ of $G$ form the vector representation of $K$ under the adjoint action. Let us denote the vector representation by $v$. Then
\begin{equation}
    [k,L^{1\mu}] = v(k)^\mu_{\ \nu} L^{1\nu}, \quad k\in\mathfrak{k}\ .
\end{equation}
Properties of invariant vector fields that we just described now directly imply
\begin{equation}
    \mathcal{L}_k \mathcal{L}_{L^{1\mu}} f^a_{\ \alpha} = (v(k)^\mu_{\ \nu}\delta^\beta_\alpha + \delta^\mu_\nu\sigma^\beta_{\ \alpha}(k))\mathcal{L}_{L^{1\nu}} f^a_{\ \beta}, \quad \mathcal{R}_k \mathcal{L}_{L^{1\mu}} f^a_{\ \alpha} = \rho^a_{\ b}(k) \mathcal{L}_{L^{1\mu}} f^b_{\ \alpha}\ . 
\end{equation}
Therefore, $\{\mathcal{L}_{L^{1\mu}}f\}$ belongs to the space $\Gamma_{\rho,\sigma\otimes v}$. Similarly, $\{\mathcal{R}_{L^{1\mu}}f\}$ is an element of the space $\Gamma_{\rho\otimes v,\sigma}$. So, Lie derivatives $\mathcal{L}_{L^{1\mu}},\mathcal{R}_{L^{1\mu}}$ are {\it shifting-operators} in the sense that they change covariance properties of $f$ in a definite way. Since invariant vector fields commute with the Laplacian, they map eigenfunctions of $\Delta$ to eigenfunctions of $\Delta$.

\vskip0.1cm By applying the radial component map to linear elements $L^{1\mu}$ in $U(\mathfrak{g})$, we get intertwiners between different Laplace operators $\Delta_{\rho,\sigma}$. For symmetric traceless tensors $\rho = (l)$, $\sigma^\ast = (l')$, the reduced Laplacian$\Delta^{(d)}_{l,l'}$ admits two weight shifting operators that change $l$ and $l'$ by one
\begin{equation}\label{shift-equations}
    \Delta^{(d)}_{l+1,l'} q_{l,l'} = q_{l,l'} \Delta^{(d)}_{l,l'}, \quad \Delta^{(d)}_{l,l'+1} \bar q_{l,l'} = \bar q_{l,l'} \Delta^{(d)}_{l,l'}\ .
\end{equation}
Note that these relations imply that if $f$ is an eigenfunction of $\Delta^{(d)}_{l,l'}$ with the eigenvalue $\lambda$, then $q_{l,l'}f$ is an eigenfunction of $\Delta^{(d)}_{l+1,l'}$ with the same eigenvalue
\begin{equation}
    \Delta^{(d)}_{l+1,l'} q_{l,l'} f = q_{l,l'} \Delta^{(d)}_{l,l'} f = \lambda q_{l,l'} f\ .
\end{equation}
An analogous statement holds for $\bar q_{l,l'}$. Explicitly, $q_{l,l'}$ and $\bar q_{l,l'}$ are given by
\begin{equation}\label{weight-shifting}
    q_{l,l'} = \partial_\theta - l \cot\theta + \frac{(y^2-1)\partial_y - l'y}{\sin\theta}, \quad \bar q_{l,l'} = \partial_\theta - l' \cot\theta + \frac{(y^2-1)\partial_y - ly}{\sin\theta}\ .
\end{equation}
One readily verifies equations \eqref{shift-equations}. To arrive at expressions \eqref{weight-shifting}, one uses the radial decompositions of $L_{1i}$ written in \eqref{radial-generators}. While the construction of operators \eqref{weight-shifting} was motivated by considerations of vector fields, their applications rely almost solely on the exchange relations \eqref{shift-equations}\footnote{Indeed, demanding the relations \eqref{shift-equations} was used to fix the final form of operators \eqref{weight-shifting}.}. In fact, the latter show that $q_{l,l'}$ is an intertwiner between "irreducible" spaces $\Gamma_{(l),(l')}\to\Gamma_{(l+1),(l')}$ rather than $\Gamma_{(l)\otimes(1),(l')}$ (and similarly for $\bar q$). This additional projection that is entailed in $q,\bar q$ is a significant property in applications.

\subsubsection{Internal weight-shifting}
\label{subsubsec: Internal weight-shifting}

We will now supplement the above operators that shift $\rho$, $\sigma$ by ones that shift $\pi$. For the above examples, this corresponds to changing $(J,q)$ while keeping $(l,l')$ constant. The fact underlying the construction of internal shifting operators is that any space $\Gamma_{\rho,\sigma}$ is a module over the zonal spherical functions $\Gamma_{0,0}$. That is, if $f_0\in\Gamma_{0,0}$ is zonal and $f^a_{\ \alpha}\in\Gamma_{\rho,\sigma}$, their product is again an element of $\Gamma_{\rho,\sigma}$
\begin{align*}
    (f_0 f)^a_{\ \alpha}(k_l g k_r) &= f_0(k_l g k_r) f^a_{\ \alpha}(k_l g k_r) = f_0(g) \rho^a_{\ b}(k_l)  f^b_{\ \alpha}(g) \sigma^\beta_{\ \alpha}(k_r) =\\
    &=\rho^a_{\ b}(k_l)  (f_0f)^b_{\ \alpha}(g) \sigma^\beta_{\ \alpha}(k_r)\ .
\end{align*}
Therefore, starting from a single function in $\Gamma_{\rho,\sigma}$, we may multiply it by zonal spherical functions to generate other elements of the same space. This is also clear by looking at matrix elements
\begin{equation}
    \pi_1(g)^a_{\ \alpha} \pi_2(g)^0_{\ 0} = \sum_{\pi\in\pi_1\otimes\pi_2} c_\pi \pi(g)^a_{\ \alpha}\ .
\end{equation}
After reduction to functions $f(\theta,y)$, we obtain relations
\begin{equation}\label{mechanism-internal-shifing}
    f^{J',0}_{0,0}(\theta) f^{J,0}_{l,l'}(\theta,y) = \sum_{\pi\in (J)\otimes (J')} c_\pi f^\pi_{l,l'}(\theta,y)\ .
\end{equation}
The first factor on the left hand side is a zonal spherical function and thus known. The second factor is also known, being obtained from the zonal spherical function $f^{J,0}_{0,0}(\theta)$ by repeated applications of external weight-shifting operators. Thus, our task is to isolate in the product the different terms appearing on the right. The set of irreducible representations $\pi$ appearing in $(J)\otimes(J')$ is of course finite and well known - we shall denote it by $\{\pi_1,\dots,\pi_n\}$. Since the functions on the right hand side of \eqref{mechanism-internal-shifing} differ only in the upper index $\pi$ and not the lower ones, they are eigenfunctions of the same differential operator $\Delta^{(d)}_{l,l'}$. Assume for the moment that all representations $\pi_i$ have different Casimir eigenvalues $C_2(\pi_i)$. If we act by $\Delta^{(d)}_{l,l'} - C_2(\pi_j)$ on the right for some fixed $j$, one term in the sum will drop out and other will get multiplied by numbers $C_2(\pi_i) - C_2(\pi_j)$. Thus, to compute a single term in the sum, we make $n-1$ subtractions, i.e.
\begin{equation}\label{internal-weight-shifting}
    f^{\pi_n}_{l,l'}(\theta,y) \sim (\Delta^{(d)}_{l,l'} - C_2(\pi_{n-1}))\dots(\Delta^{(d)}_{l,l'} - C_2(\pi_1))\left(f^{J',0}_{0,0}(\theta) f^{J,0}_{l,l'}(\theta,y) \right)\ .
\end{equation}
If some representations from $\{\pi_1,\dots,\pi_n\}$ are not distinguished by the values of the quadratic Casimir $C_2$, one should also apply similar projections with higher Casimirs (in practise, we generally do not need to do this - see however discussions below, especially in $d=5$). Notice that both external \eqref{weight-shifting} and internal \eqref{internal-weight-shifting} weight-shifting operators require only are expressed in terms of reduced operators only (i.e. depending only on $\theta$ and $y$). Using them, all spherical functions are computed by the sequence of steps that we shall now detail.

\subsection{Solution theory}
\label{solution theory STT-STT}
In this subsection, we explain how to use the previously define tools to obtain a general solution to the Laplacian.

\subsubsection{Zonal spherical functions}

As we have said, zonal spherical functions are matrix elements with trivial $\rho$ and $\sigma$. They are obtained by solving the differential equation
\begin{equation}\label{1-sided}
    \Delta^{(d)}_{0,0} f = - J(J+d-3) f, \qquad \Delta^{(d)}_{0,0} = (1-t^2)\partial_t^2 - (d-2)t\partial_t \ .
\end{equation}
We have written $t=\cos\theta$. The solutions with correct boundary conditions are Gegenbauer polynomials
\begin{equation}\label{zonal-spherical}
   f^{J,0}_{0,0} = C^{(\frac{d-3}{2})}_J(t)\ .
\end{equation}
These agree with the scalar partial waves, as mentioned already in \eqref{zonal-section2}. In four and five dimensions, they reduce Legendre and Chebyshev polynomials
\begin{equation}
    d=4:\quad f^{J,0}_{0,0} = P_J(t), \qquad d=5: \quad f^{J,0}_{0,0} = U_J(t) = \frac{\sin((n+1)\theta)}{\sin\theta}\ .
\end{equation}

\subsubsection{Higher solutions}

Let us now describe the method to obtain admissible eigenfunctions of the Laplacian \eqref{final}\footnote{There is a distinction between eigenfunctions of \eqref{final} and matrix elements - namely, $\Delta^{(d)}_{l,l'}$ admits eigenfunctions which do not correspond to any matrix elements. We reserve the adjective 'admissible' for the ones that do.}.\\
While we focus on the case of external symmetric traceless representations, the general idea is the same in more involved cases as well - we give details of the procedure for MST-STT and MST-MST systems in Appendix \ref{External-mixed-symmetry-tensors}. One always starts from certain solvable {\it boundary cases}, characterised by a very particular choice of quantum numbers, and shows that any admissible solution is obtained from them by acting with weight-shifting operators. The steps are summarised by the following diagram and explained in more details in the next paragraph,

\begin{center}
\begin{tikzpicture}
    \node (f) at (0,0) {$f_{0,0}^{J,0}$};
    \node (g) at (3,0) {$f_{l,0}^{J,0}$};
    \node (h) at (6,0) {$f_{l,l'}^{J,0}$};
    \node (i) at (9,0) {$f_{l,l'}^{J,q} \, .$};
    
    \draw[->] (f) -- node[above] {$(q)^l$} (g);
    \draw[->] (g) -- node[above] {$(\bar{q})^{l'}$} (h);
    \draw[->] (h) -- node[above] {internal} (i);
    \draw[->] (h) -- node[below] {w.s.} (i);
\end{tikzpicture}
\end{center}

\paragraph{Strategy}

The process of constructing solutions is divided in three steps:\footnote{Here we describe the procedure for $d>5$. The case $d=5$ is describe in Section \ref{Exceptions in five dimensions}.}
\begin{enumerate}
\item The starting point is the zonal spherical function, the eigenfunction of $\Delta_{0,0}^{(d)}$ with the eigenvalue $C_2((J,0))$, computed in the previous subsection, \eqref{zonal-spherical}.

\item The next step involves the external weight-shifting. We wish to arrive to the eigenfunction of $\Delta_{l,l'}^{(d)}$ with the eigenvalue $C_2((J,0))$. This is done by repeatedly acting with $q$ or $\bar{q}$ until we reach the desired values of $l,l'$. 

\item Finally, we use the internal weight-shifting to get the solution we want, the eigenfunction of $\Delta_{l,l'}^{(d)}$ with the eigenvalue $C_2((J,q))$. The result of the previous step is multiplied by the zonal spherical function $f^{q,0}_{0,0}$, where we have put the index $q$ because we need to move our solution 'by $q$ steps' to go from $(J,0)$ to $(J,q)$. After the multiplication, one gets a linear combinations of solutions corresponding to all irreducible representations in the tensor decomposition of $(J,0) \otimes (q)$, one of which is $(J,q)$. We get rid of the other representations $\pi_i$ by projecting them out with $\Delta^{(d)}_{l,l'} - C_2(\pi_i)$, where $C_2(\pi_i)$ is the Casimir value in the representation $\pi_i$. 
\end{enumerate}

The {\it Mathematica} function that carries out the above procedure is \\$\texttt{Generate\$STTSTT\$yBasis[d,l,l',J,q][t,y]}$.

\paragraph{Example}
Let us now get for instance the solution to $\Delta^{(d=7)}_{2,1}$ with the eigenvalue $C_2((3,1))$ and $d>5$, following the described procedure. We start from the zonal with $J=3$, $f^{3,0}_{0,0}=4t(8 t^2 -3)$, and then act with external weight shifting operators to arrive to $l=2$ and $l'=1$, e.g.
\begin{equation}
   f^{3,0}_{2,1}(\theta)= \bar{q}_{2,0}(q_{1,0}(q_{0,0}(f^{3,0}_{0,0}(\theta))))\propto (1 - t^2)^{\frac{1}{2}} (3 t^2 - 2 t y-1) \, .
\end{equation}
In the previous equation we drop the numerical prefactor (which is not important since the function will be normalised only at the very end) and we put the proportionality sign $\propto$. The next step is to change the internal representation $(3,0) \to (3,1)$. We need to shift $q$ by one and so we multiply $f^{3,0}_{2,1}(t,y)$ by $f^{1,0}_{0,0}(t)$. The relevant tensor product decomposition reads
\begin{equation}
    (3,0) \otimes (1,0) = (2,0) \oplus (3,1) \oplus (4,0) \, .
\end{equation}
The last thing we need to do is to project to the internal representation we want, that is $(3,1)$,
\begin{equation}
    f^{3,1}_{2,1}(t,y) = (\Delta^{(d)}_{2,1} - C_2(2,0))(\Delta^{(d)}_{2,1} - C_2(4,0))\left(f^{3,0}_{2,1}(t,y) f^{1,0}_{0,0}(t) \right)\propto (1 - t^2) (6 t^2 y -5t-y)  \ .
\end{equation}

\subsection{Matrix elements in the Gelfand-Tsetlin basis and checks}
\label{subsec: Matrix elements in the Gelfand-Tsetlin basis and checks}

The procedure we have given computes non-normalised matrix elements in the function space basis for carrier spaces of $\rho$ and $\sigma$, \eqref{differential-basis-1}-\eqref{differential-basis-2}. To normalise them, it is useful to go to the Gelfand-Tsetlin (GT) basis, which is often used in the literature. Furthermore, using the GT basis allows to compare a subset of our matrix elements to known results, \cite{Vilenkin:1993:RLG2}. Here we only state how to obtain normalised matrix elements \eqref{matrix-elements-final} from the function $f(\theta,y)$. More details and the general picture are given in Appendix \ref{From polynomials to the Gelfand-Tsetlin basis}.

\paragraph{To Gelfand-Tsetlin basis} Fix $l$ and $l'$. Given a function $f^{J,q}_{l,l'}(\theta,y) = a_0(\theta) + a_1(\theta)y + \dots$, representing $\pi_{J,q}$ in the function space basis, the corresponding matrix elements in the Gelfand-Tsetlin basis $F^{J,q}_{l,l',j}(\theta)$ are obtain from the following relation,
\begin{align}
    f^{J,q}_{l,l'}(\theta,y)= \sum_{j} F^{J,q}_{l,l',j}(\theta) C^{\left(\frac{d-5}{2}\right)}_{j}(y) \,,
\end{align}
where $j$ runs from 0 to $\text{min}(l,l')$ for $d>5$\footnote{The case $d=5$ is describe in Section \ref{Exceptions in five dimensions}.}. The {\it Mathematica} command that gives the map fro the change of basis is $\texttt{toGTbasis\$STTSTT[d][y][f]}$, more details in Appendix \ref{Practical implementation of the algorithm}.

\paragraph{Normalisation}
We compute the normalisation of the matrix elements using orthonormality of the Gelfand-Tsetlin basis and the action of infinitesimal generators defined in \eqref{Action-infinitesimal-generators}. We expand $F^{J,q}_{l,l',j}(\theta)$ in $\theta$, and we impose that the $|l-l'|$-th coefficient of the expansion has to be equal to $ \langle l,j|L^{|l-l'|}_{12} | l',j\rangle$.
The {\it Mathematica} command to normalise is $\texttt{normaliseSTTSTTinGT[d,l,l',J,q][$\theta$][f]}$, and the one that gives the normalised matrix element in the GT basis is $\texttt{Generate\$STTSTT\$GTBasis[d,l,l',J,q][$\theta$]}$, see Appendix \ref{Practical implementation of the algorithm}.

\paragraph{Check}Matrix elements for external and internal symmetric traceless tensor, i.e. for $q=0$, admit the integral representation, \cite{Vilenkin:1993:RLG2}, Section 9.5,
{\small\begin{align}\label{integral-rep}
    & F^{d-1,J}_{l,l',j}(\theta) = \\
    & B^{d-1,J}_{l,l',j} \int_0^\pi d\phi\ (\sin\phi)^{2j+d-4}(\cos\theta-i\cos\phi\sin\theta)^{J-j} C^{(j+\frac{d-4}{2})}_{l-j}(\cos\phi) C^{(j+\frac{d-4}{2})}_{l'-j}\left(\frac{\cos\phi\cos\theta-i\sin\theta}{\cos\theta-i\cos\phi\sin\theta}\right)\,,\nonumber
\end{align}}

where
{\small\begin{align*}
    B^{d-1,J}_{l,l',j} &=i^{l-l'}\sqrt{\frac{(J+l+d-4)!(J-l)!}{(J+l'+d-4)!(J-l')!}} \frac{2^{2j+d-6}\Gamma\left(j+\frac{d-4}{2}\right)^2}{\pi}\times\\ &\times\sqrt{\frac{(l-j)!(l'-j)!(2l+d-4)(2l'+d-4)}{(l+j+d-5)!(l'+j+d-5)!}}\ .
\end{align*}}

We checked our results for $q=0$ exchange against \eqref{integral-rep} in a number of examples and observed perfect agreement. It might be worth mentioning that, at least for {\it Mathematica}, doing the integral is considerably more time-consuming than constructing solutions through weight-shifting. For $q\neq0$, our method goes beyond \cite{Vilenkin:1993:RLG2}.

\paragraph{Remark: conformal to $S$-matrix limit} The framework of spherical functions also applies to conformal partial waves. Indeed, the latter may be viewed as spherical functions for a rank-two Gelfand pair, with $G=SO(d+1,1)$ being the conformal group and $K = SO(1,1)\times SO(d)$ the group of dilations and rotations, \cite{Schomerus:2016epl,Buric:2022ucg}. The reduced Laplacian acting on these functions reads, \cite{Buric:2022ucg},
\begin{small}
    \begin{align}
     H &= \partial_{t_1}^2 + \partial_{t_2}^2 +\frac{1 - D'^2_+ + 2\cosh(t_1+t_2)
     D'_+D_+ -  D^2_+}{2\sinh^2(t_1+t_2)}
     + \frac{1 - D'^2_- + 2\cosh(t_1-t_2)D'_-D_- - D^2_-}{2\sinh^2(t_1-t_2)}\nonumber\\[2mm]
       & + \frac{M'_{2a}M'_{2a}-2\cosh t_1 M'_{2a}M_{2a}+M_{2a}M_{2a}-\frac14(d-2)(d-4)}{\sinh^2 t_1}
       \label{universal-Hamiltonian}\\[2mm]
       & + \frac{M'_{3a}M'_{3a}-2\cosh t_2 M'_{3a}M_{3a}+M_{3a}M_{3a} - \frac14 (d-2)(d-4)}{\sinh^2 t_2}-
       \frac12 L^{ab}L_{ab}- \frac{d^2-2d+2}{2}\ .\nonumber
\end{align}
\end{small}

Here, $D$ and $M_{ij}$ are the usual dilation and rotation generators of the conformal algebra and $D_\pm = D \pm i M_{23}$ (for the range of indices and commutation relations, see \cite{Buric:2022ucg}). The objects $D_\pm,M_{ij}$ and $D'_\pm,M'_{ij}$ in \eqref{universal-Hamiltonian} are to be interpreted as operators representing abstract generators $D_\pm,M_{ij}$ in two representations of $K$. The latter are essentially the tensor products of field representations at points $\{3,4\}$ and $\{1,2\}$, respectively.\footnote{This is a somewhat imprecise description. Details about the two representations of $K$ are given in \cite{Buric:2022ucg}.} On the other hand, the two variables $t_1$, $t_2$ are closely related to the radial coordinates of \cite{Hogervorst:2013sma}
\begin{equation}\label{radial-variables}
     r = -e^{-t_2}, \quad \eta = \cos(i t_1)\ .
\end{equation}
The $S$-matrix limit is given by $r\to0$, i.e. $t_2\to\infty$ and relates conformal correlators in $d$ dimensions to amplitudes in $d+1$ dimensions. Indeed, taking this limit in \eqref{universal-Hamiltonian} and acting on functions of the form $f(t_1)e^{\Delta t_2}$, we get
\begin{small}
    \begin{align*}
     H &= - \partial_\theta^2 - \frac{M'_{2a}M'_{2a}-2\cos\theta M'_{2a}M_{2a}+M_{2a}M_{2a}-\frac14(d-2)(d-4)}{\sin^2 \theta}+\\ &- \frac12 L^{ab}L_{ab}- \frac{d^2-2d+2}{2}+\Delta^2\ .
\end{align*}
\end{small}

We have introduced $\theta = it_1$, in accord with \eqref{radial-variables}. The last operator is, up to the conventional overall minus sign and an additive constant, the Schr\"odinger form of the Laplacian \eqref{main} in $d+1$ dimensions (by the "Schr\"odinger form of $\Delta$", we mean the unique operator related to $\Delta$ by conjugation, $\omega(\theta)\Delta\omega(\theta)^{-1}$, that does not contain the term linear in the derivative $\partial_\theta$). The representations $\rho$ and $\sigma^\ast$ are restrictions to $SO(d-1)$ of $\rho_1\otimes\rho_2^w$ and $\rho_3\otimes\rho_4^w$, where $\rho_i$ are the representations of $SO(d)$ characterising the four conformal fields that enter the correlator.

\section{Exceptions in five dimensions}
\label{Exceptions in five dimensions}

The general theory we developed has to be adjusted if the spacetime dimension is low. The critical dimension for which the theory becomes generic depends on the depth, i.e. the number of non-trivial GT labels, of representations $\rho$ and $\sigma$. When $\rho$ and $\sigma$ are symmetric traceless tensors, the theory as written in the last section applies for $d>5$. Since the matrix elements for $d=4$ were reviewed in Section \ref{Illustration Wigner d}, it remains to treat the case $d=5$. This is the purpose of the present section.

In five spacetime dimensions, the groups entering the discussion are
\begin{equation}
    G = SO(4), \quad K = SO(3), \quad M = SO(2)\ .
\end{equation}
As before, we will be imprecise about the global structure of the groups. The reader can assume that we are working with universal covering groups $\text{Spin}(4)$ and $SU(2)$ of the above, or equivalently with projective representations. Representations of $G$ will be labelled either by Gelfand-Tsetlin labels $(J,q)$ or by two $SU(2)$-spins $[j_1,j_2]$. The relation between two sets of labels reads
\begin{equation}
    J = j_1 + j_2, \quad q = j_1 - j_2\ .
\end{equation}
In the remainder of the section, we will compute spherical functions on $G$ with definite covariance properties under left and right actions of $K$.

\subsection{Enumeration of matrix elements}

Let as before $\rho = (l)$, $\sigma = (l')$ and $\pi = (J,q)$. Contrary to the higher-dimensional cases, the label $q$ is allowed to be negative, with $|q|\leq J$. Representations $\rho$ and $\sigma$ appear in the restriction of $\pi$ to $K$ if and only if
\begin{equation}\label{selection-rules-d=5}
    J \geq l,l' \geq |q|\ .
\end{equation}
The Lie algebra $\mathfrak{g}$ is not simple and has two independent quadratic Casimir elements. We denote by $C_2^{(i)}$, $i=1,2$, the Casimirs of two commuting $\mathfrak{su}(2)$ subalgebras. These are normalised to take values $-2j_i(j_i+1)$ in the representation $[j_1,j_2]$. We denote $C_2 = C_2^{(1)} + C_2^{(2)}$ and $\tilde C_2 = C_2^{(1)} - C_2^{(2)}$. With these conventions,
\begin{equation}
    C_2(\pi_{J,q}) = -J(J+2) - q^2\,, \quad \tilde C_2(\pi_{J,q}) = -2q (J+1)\,, \label{eq: Casimirs in 5d}
\end{equation}
so, in particular, the normalisation of $C_2$ is coherent with \eqref{Casimir-values}.
\smallskip

Now assume that $l$, $l'$, $J$ and $q$ are fixed. Vectors $e^a$ and $e_\alpha$ may be written as
\begin{equation*}
    e^a = |J,q;l;j\rangle, \quad e_\alpha = |J,q;l';j'\rangle\,,
\end{equation*}
where we have separated rows of the Gelfand-Tsetlin patterns by semicolons. Unlike in higher dimensions, $j$ can be negative - it runs over $\{-l,-l+1,\dots,l\}$. Non-zero matrix elements have $j + j' = 0$ and we put
\begin{equation}\label{independent-matrix-elements-d=5}
    F^{4,J,q}_{l,l',j}(\theta) = \langle l,j|\pi_{J,q}(e^{\theta L_{12}})| l',-j\rangle\ .
\end{equation}
This is the analogue of the set of higher-dimensional independent matrix elements \eqref{matrix-elements-final}. Due to the modified range for $j$, the number of independent functions is now
\begin{equation}\label{number-of-matrix-element-d=5}
    N(l,l') = 2\text{min}(l,l') + 1\ .
\end{equation}

\subsection{Reduced Casimir operators and solution theory}

The function space realisation of spin-$l$ representation of $K$ reads
\begin{small}
    \begin{equation}\label{function-space-realisation-d=5}
    \rho(L_{23}) = -i(x\partial_x - l),\quad \rho(L_{24}) = -\frac12 \Big((1+x^2)\partial_x - 2 l x \Big), \quad \rho(L_{34}) = \frac{i}{2}\Big((1-x^2)\partial_x + 2 l x\Big)\ .
\end{equation}
\end{small}

These operators act on the space of polynomials in $x$ of degree less than or equal to $2l$. Similarly, representation $\sigma^\ast$ is realised in terms of operators in coordinates $x'$ and has spin $l'$. These expressions are special cases of the general ones written above when the index $a$ runs over the one-element set $\{4\}$. The same is true for solutions to $M$-invariance conditions \eqref{m-invariance}, which still take the form \eqref{form-of-functions-STT}. We again denote the single invariant variable by $y$. The reduced Laplacian is given by \eqref{final}. However, now the space on which it is to be considered is that of functions of $\theta$ and $y$, whose dependence on $y$ is of the form
\begin{equation}\label{space-of-funtions-d=5}
    P(y) + Q(y)\sqrt{y^2-1}, \quad \text{with}\quad \text{deg}(P)\leq \text{min}(l,l'), \quad \text{deg}(Q)\leq \text{min}(l,l') - 1\ .
\end{equation}
Here, $P$ and $Q$ are polynomials. We see that $\Delta^{(5)}_{l,l'}$ preserves this space. The other Casimir reduces to
\begin{equation}
    \tilde\Delta_{l,l'} = \partial_\theta\partial_\phi + \frac{\sin\phi\partial_\phi^2 + (\cos\theta - (l+l')\cos\phi)\partial_\phi - l l'\sin\phi}{\sin\theta}\ .
\end{equation}
We have displayed the second Casimir using the variable $\phi$ related to $y$ by $y = \sin\phi$. It appears that expressed in this variable $\tilde\Delta_{l,l'}$ takes the simplest form.
\smallskip

Since the reduced Laplacian in $d=5$ coincides with its higher-dimensional counterpart, it also admits external weight-shifting operators \eqref{weight-shifting} satisfying the exchange relations \eqref{shift-equations}. The principal difference in the solution theory compared to previous cases follows from the fact that there are many pairs of representations $\pi$ of $G$ with equal values for the Casimir $C_2$. Namely, $C_2(\pi_{J,q}) = C_2(\pi_{J,-q})$. We account for this by using both $\Delta_{l,l'}$ and $\tilde\Delta_{l,l'}$ to project out representations in the algorithm described in Section \ref{solution theory STT-STT}. The implementation in {\it Mathematica} of the algorithm is given by the function $\texttt{Generate\$STTSTT\$yBasis[5,l,l',J,q][t,y]}$, more details are given in Appendix \ref{Practical implementation of the algorithm}.

\subsection{Eigenfunctions in the Gelfand-Tsetlin basis}

To express spherical functions in the GT basis, the only thing we need is write the usual vectors $|l,j\rangle$ of an $SO(3)$-irreducible representation in the realisation \eqref{function-space-realisation-d=5}. The relation between the two reads
\begin{equation}\label{patterns-functions-d=5}
  |l,j\rangle =  c_{l,j}\left(\frac{1+x}{1-x}\right)^j\,, \quad c_{l,j} = \frac{1}{i^j \sqrt{2^l (l+j)!(l-j)!}}
\end{equation}
Given a function $f(\theta,y) = a_n(\theta)y^n + i b_m(\theta)y^m\sqrt{y^2-1}$ that represents the matrix elements $\pi_{J,q}$ in the differential basis, we expand it as
\begin{equation}
    f(\theta,y) = K_{mn}(\theta)\chi^m \chi'^n, \qquad \chi = \frac{1+x}{1-x}, \quad \chi' = \frac{1+x'}{1-x'}\ .
\end{equation}
The corresponding matrix elements in the Gelfand-Tsetlin basis read
\begin{equation}\label{function-to-GT-d=5}
   F^{4,J,q}_{l,l',j}(\theta) = c_{l,j} c_{l',j} K_{jj}(\theta)\ .
\end{equation}
The {\it Mathematica} function that gives matrix elements in the GT basis is given in \eqref{GenerateFunctions-nb}. Compared to \eqref{function-to-GT-d=5}, the function includes a further normalisation, see Appendix \ref{From polynomials to the Gelfand-Tsetlin basis}. We have checked our results against integral representations from \cite{Vilenkin} in a number of cases where the comparison is possible and observed perfect agreement.

\subsection{An example: photon scattering in five dimensions}

In this subsection, we illustrate the theory developed in previous sections on the example of photon scattering in $d=5$ dimensions. Our results for this case agree with \cite{correspondence}.
\smallskip

As we mentioned, in order to define partial waves, a choice is needed for spaces of three- and four-point tensor structures. Let us adopt the notation of \cite{Caron-Huot:2022jli} for this and the following paragraph. Then, three-point structures are encoded by vertices $v^j(n,e_1,e_2)$, which are are polynomials in components of vectors $e_1$, $e_2$, $n$, $w_1$ and $w_2$. The first three vectors are constructed out of momenta $p_i$ and polarisations $\epsilon_i$, whereas the last two characterise the spin of the exchanged particle (the authors of \cite{Caron-Huot:2022jli} give vertices relevant for scattering of gravitons. In the case of photons, they provided us with their choice of vertices to allow for comparison). On the other hand, four-point tensor structures may be chosen as\footnote{We use the same definition and notation of \cite{Chowdhury:2019kaq}.}
\begin{equation}\label{photon-tensor-structures-DSD}
    H_{14} H_{23}, \quad H_{13} H_{24}, \quad H_{12} H_{34}, \quad X_{1243}, \quad X_{1234}, \quad X_{1324}, \quad S\ .
\end{equation}
These four-point structures are referred to as generators of the 'local module', in the terminology of \cite{Chowdhury:2019kaq} (where they are contrasted with the generators of the 'bare module'). We shall denote these structures by $T^I$, $I=1,\dots,7$. They are polynomials in momenta $p_i$ and polarisations $\epsilon_i$ of the four particles. Given an internal representation $\pi$ and a pair of vertices $v^i,v^j$, the partial wave $g^{ij}_\pi$ can be expanded in the above structures and is thus encoded in seven scalar functions $g^{ij}_{\pi I}$ of the scattering angle $\theta$. In conclusion, the partial waves are given by the data $\{v^i,T^I,g^{ij}_{\pi I}(\theta)\}$.
\smallskip

Now we revert back to our notation. Since we use $w_i$ to denote spin states of external particles, the $w_i$ of \cite{Caron-Huot:2022jli} shall be renamed to $r_i$. To compare the above to our partial waves, we first fix particles to the frame as explained in Section \ref{Tree-point structures}
\begin{align}
    & p_1 = \frac{\sqrt{s}}{2}(1,1,0,0,0),  \qquad p_3 = -\frac{\sqrt{s}}{2}(1,\cos\theta,\sin\theta,0,0)\,,\label{momentum-frame-1}\\
    & p_2 = \frac{\sqrt{s}}{2}(1,-1,0,0,0), \quad\ p_4 = -\frac{\sqrt{s}}{2}(1,-\cos\theta,-\sin\theta,0,0)\ .\label{momentum-frame-2}
\end{align}
Secondly, polarisations $\epsilon_i$ are expressed in terms of spin vectors $w_i$ by $\epsilon_i = \Lambda_{p_i}(0,0,w_i)$. Recall from Section \ref{Single particle states}, where this relation was established, that $w_i$ is a vector in the spin-1 representation of $SO(3)$, and $\Lambda_{p_i}$ is the standard boost \eqref{standard-boosts-massless}. By these substitutions, vertices $v^i(n,e_1,e_2)$ are turned into polynomials in components of $w_1$, $w_2$, $r_1$ and $r_2$, $SO(3)$-invariant elements of $(1)\otimes(1)\otimes\pi$. Further, the structures \eqref{photon-tensor-structures-DSD} become elements of the carrier space of $(1)^{\otimes4}$ of $SO(3)$, whose coefficients are functions of $\theta$, namely
\begin{equation*}
    T^I \to T^I_{\alpha\beta\gamma\delta}(\theta) w_1^\alpha w_2^\beta w_3^\gamma w_4^\delta = T^I_{\rho\mu\sigma\nu}(\theta) w_{12}^{\rho\mu}w_{34}^{\sigma\nu}, \qquad I=1,\dots 7\, \quad \alpha,\beta,\gamma,\delta = 2,3,4\ .
\end{equation*}
In the second equality, we used the $SO(3)$ Clebsch-Gordan coefficients to go from the tensor product basis of $(1)\otimes(1) = (0)\oplus(1)\oplus(2)$ to the irreducible-components basis. The same was done for pairs of particles $\{1,2\}$ and $\{3,4\}$. In the usual $SO(3)$ notation
\begin{equation*}
    w^{\rho\mu} = |j,m\rangle, \quad \text{with} \quad \rho = (j), \quad \mu = m, \quad j = 0,1,2, \quad m = -j,\dots,j\ .
\end{equation*}
To compare $g^{ij}_{\pi I}$ to matrix elements $\pi(e^{\theta L_{12}})^{\rho\mu}{}_{\sigma\nu}$, one is required to relate vertices $v^i$ in the frame to \eqref{frame-3pt}. Consider for concreteness the symmetric traceless exchange. Then $r_2=0$ and we write $r_1 = r$. The vertices $v^i$ of \cite{correspondence} (first three of them) become
\begin{align}\label{vertices-1}
    & v^0 \to \frac{k_J}{\sqrt{3}} r_1^J \left(w_1^2 w_2^2 + w_1^3 w_2^3 + w_1^4 w_2^4\right)\,,\\
    & v^2 \to k_J\sqrt{\frac{J(J-1)}{6(J+2)(J+3)}} r_1^J \left(-2 w_1^2 w_2^2 + w_1^3 w_2^3 + w_1^4 w_2^4\right),\\
    & v^1 \to i k_J \sqrt{\frac{J}{2(J+2)}} r_1^J (w_1^4 w_2^3 - w_1^3 w_2^4)\ .\label{vertices-3}
\end{align}
The following remark is in order. We first verify that $v^i$ obtained by substitutions are $SO(3)$-invariant. Then without loss of generality we display results 'in the frame' $r = (r_1,r_2,0,0)$. It is immediately verified that \eqref{vertices-1}-\eqref{vertices-3} project to $(0)$, $(2)$ and $(1)$ external subspaces, respectively. Therefore, vertices $v^i$ agree precisely with our tensor structures \eqref{frame-3pt}, up to normalisation. The same comments apply to all other vertices from \cite{correspondence}. Therefore, our theory should imply the relation\footnote{Recall from Appendix \ref{appendix:discrete symmetry} that $\Pi^{\rho\mu}{}_{\sigma\nu} = \pi^{\rho\mu}{}_{\sigma\nu} + (\pi^\ast)^{\rho\mu}{}_{\sigma\nu}$, where $(J,q)^\ast = (J,-q)$.}
\begin{equation}
    g^{\rho\sigma}_{\pi I}(\theta) T^I_{\rho\mu\sigma\nu}(\theta) = C(\rho,\sigma,\pi) \Pi(e^{\theta L_{12}})^{\rho\mu}{}_{\sigma\nu}\,,
\end{equation}
and the latter was explicitly verified. The coefficients $C(\rho,\sigma,\pi)$ are obtained by comparison of normalisations of vertices in the two approaches.

\section{Summary and outlook}
\label{Summary and outlook}
In our study, we have developed a comprehensive approach to compute partial waves for $2\to2$ scattering involving spinning particles. Our findings have broad applicability to various processes with four particles, whether they are massive or massless, of spins $J_1,\dots,J_4$. Importantly, this approach is applicable in any spacetime dimension $d$. After establishing a connection between partial waves and specific matrix elements of the rotation group $SO(d-1)$, we employed a novel algebra of weight-shifting operators to compute the latter. We have included a valuable {\it Mathematica} code that effectively implements our method and generates the aforementioned matrix elements. In the remaining few paragraphs, we discuss some of the possible applications and extensions of our results.
\medskip

The identification of partial waves with spherical functions may lead to new insights into analytic properties of the S-matrix. Among the goals in this direction would be a spinning generalisation of the Froissart-Gribov formula \cite{Gribov:1961ex,Correia:2020xtr} as previously done in four dimensions, see for instance \cite{Martin:1970hmp}. It is an appealing idea that the 'Q-functions' appearing in this tentative formula are obtained from the scalar Q-functions by the same sequence of weight-shifting operators that produces spinning waves from the scalar ones. Another regime for which the known results about spherical functions and their asymptotic properties may have implications is the Regge behaviour of spinning amplitudes, \cite{Haring:2022cyf}.

More directly, our results allow to bootstrap previously unavailable systems, for example the scattering of particles of different type. While our focus has been on bosonic matrix elements, it is noteworthy that these are the only ones that appear as partial waves in the scattering of four fermions. On the other hand, the computation of fermionic matrix elements that would appear in scattering involving both bosons and fermions is certainly within reach of our methods. Since we treat particles of arbitrary spin, another natural avenue for applications are theories involving strings \cite{Guerrieri:2021ivu,Aharony:1999ks}, or string-like objects. Perhaps more interestingly, it is noted that partial waves for spinning $2\to2$ scattering are sufficient to construct multi-particle partial waves for scalars - the latter are simply the products of the former (see \cite{Costa:2023wfz} for a recent review and references to original literature). Therefore, the present work paves the way for analysis of processes involving more than four particles.
\medskip

There are several directions of generalisation of the above constructions. A very natural completion would be to study massless exchanges, in analogy to what has been done for conserved currents in the CFT bootstrap. At present, these exchanges are usually forbidden by the assumptions made on the S-matrix, but they may become more relevant in future studies. Another extension is to consider supersymmetric theories. The supersymmetric waves are expected to admit a description within harmonic analysis, similarly to superconformal blocks \cite{Buric:2019rms}. Spherical functions on supergroups in general carry an action of additional invariant differential operators. It is an interesting question to determine how these enrich the algebra of Laplacians and weight-shifting operators. Finally, the framework of spherical functions is by no means limited to $S$-matrix partial waves and extends to conformal blocks and specific classes of defect-channel and bulk-channel conformal blocks in dCFTs. In all these cases, there exists a universality in spin, enabling the construction of the weight-shifting algebra. We believe that many other physical systems of interest can be encompassed by this framework, including certain de Sitter/cosmological correlators \cite{Baumann:2019oyu,Baumann:2020dch}. Determining whether a given problem is described by spherical functions is a straightforward kinematical question.
\medskip

Consequently, we conclude with a brief discussion of the place of our constructions within the broader context of representation theory. The provided code represents the first systematic implementation of the idea proposed in \cite{Buric:2022ucg} for computing spherical functions through an algebra of differential operators obtained via Harish-Chandra's radial component map, specifically ${\Delta,q,\bar q}$. Our analysis focused on rank-one systems with a compact group $G$, however these restrictions can be relaxed without significant difficulty.
\smallskip

As we have mentioned on multiple occasions, the radial component map, which computes the reduction of differential operators to spaces of spherical functions, is defined in the general context of Gelfand pairs $(G,K)$. Zonal spherical functions within this more general framework are (up to a prefactor) the Heckman-Opdam hypergeometric functions associated with root systems \cite{Heckman-Opdam}, and they have been extensively studied. The results from \cite{Buric:2022ucg} for rank-two Gelfand pairs indicate that the construction of the weight-shifting algebra can be carried out independently of the rank in essentially the same way. However, the corresponding spinning higher-rank spherical functions remain largely unexplored, except for the rank-two case where examples from the CFT literature provide valuable insights, \cite{Costa:2011dw,Karateev:2017jgd}. Furthermore, the transition from compact to non-compact groups does not appear to present significant difficulties and is mainly reflected in the set of zonal spherical functions \cite{Buric:2022ucg}. It might be worth mentioning that the spherical functions we computed in this study, in the special cases where $\rho=\sigma$, correspond to matrix hypergeometric functions (see \cite{Koelink_2012,Tirao_2014} and references therein). We believe that the presented method and code can be useful resources for researchers studying these special functions.
\smallskip

In addition to viewing weight-shifting as an efficient computational tool, it is intriguing to investigate the algebraic structure formed by these operators in greater detail. From a speculative standpoint, one might investigate the possibility of finding a set of operators that would replace the internal weight-shifting procedure with a more efficient alternative. The resulting algebra, containing Laplacians, external shifting operators, and these new internal shifting operators, would be of considerable importance for the representation theory of $(G,K)$. With this goal in mind, it could be useful to view the Laplacians $\Delta_{\rho,\sigma}$ as spinning Calogero-Sutherland Hamiltonians, \cite{Buric:2022ucg}, and search for potential dual models of Ruijsenaars-Schneider type \cite{Arutyunov:2019wuv,Arutyunov:2019alw}. However, this program is still in very early stages and requires further investigation.
\bigskip

\paragraph{Acknowledgements:} We wish to thank Yue-Zhou Li for sharing unpublished results for purposes of comparison and Francesco Bertucci, Johan Henriksson, Brian McPeak and Volker Schomerus for useful discussions. This work received funding from a research grant under the project H2020 ERC STG 2017 G.A. 758903 "CFT-MAP".

\appendix

\section{Some group theory background}
\label{Some group theory background}

In this appendix, we give a review of certain statements and constructions from representation theory that are used in the main text. For more details, the reader is referred to e.g. \cite{Kirillov,Vilenkin}.

\subsection{Spherical functions}

Let $G$ be a simple Lie group and $K\subset G$ a Lie subgroup. Given two finite-dimensional representations $\rho,\sigma$ of $K$, let $\Gamma_{\rho,\sigma}$ be the space of vector-valued functions on $G$
\begin{equation}\label{spherical-functions-general}
    \Gamma_{\rho,\sigma} = \{f: G \to \text{Hom}(W_r,W_l)\ | f(k_l g k_r) = \rho(k_l) f(g) \sigma(k_r)\}, \qquad k_l,k_r\in K,\ g\in G\ .
\end{equation}
Here, $W_l$ and $W_r$ are the carrier spaces of $\rho$ and $\sigma$, respectively. It can be shown that the space $\Gamma_{0,0}$ specified by trivial representations $\rho$ and $\sigma$ is closed under convolutions. If this convolution algebra is commutative, $K$ is said to be a spherical subgroup of $G$ and $(G,K)$ is called a Gelfand pair. There are several criteria for determining if two groups $G,K$ form a Gelfand pair, \cite{Kirillov}. A particularly simple one states, under appropriate assumptions on $G$ and $K$: if $\theta$ is an involutive automorphism of $G$ and $K$ its fixed point set, then $(G,K)$ is a Gelfand pair.

\smallskip

Given a Gelfand pair $(G,K)$, the group $G$ admits, at least locally, a Cartan decomposition
\begin{equation}\label{Cartan-decomposition-general}
     G = K A_p K\ .
\end{equation}
By this we mean that almost all elements of $G$\footnote{Up to a subset of measure zero.} may be factorised as $g = k_l a k_r$ with $k_{l,r}\in K$ and $a\in A_p$. Here, $A_p\subset G$ is a certain abelian group, whose dimension is called the real (or spilt) rank of $(G,K)$ and denoted rank$(G,K)$. Due to covariance properties \eqref{spherical-functions-general}, spherical functions $f\in\Gamma_{\rho,\sigma}$ are uniquely determined by their restrictions to the subgroup $A_p$ - for this reason, they may be regarded as vector-valued functions of rank$(G,K)$ variables.

\subsection{Gelfand-Tsetlin formulas}

Besides being a useful labelling scheme for states, the Gelfand-Tsetlin patterns allow for an explicit description of the action of Lie algebra generators in any irreducible representations of $SO(n)$, \cite{Gelfand-Tsetlin}. Since we use these GT formulas to normalise matrix elements, we shall spell them out, following \cite{vilenkin1992representation}, Chapter 18.1.2. Adapted to our conventions, and putting $n=d-1$, we have
\begin{align}\label{Action-infinitesimal-generators}
    L_{d-p,d-p+1} v(M) &= \sum_j A^j_p(M) v(M^{+j}_{p}) - \sum_j A^j_p(M^{-j}_p) v(M^{-j}_p)\ .
\end{align}
Here $v(M)$ denotes a GT pattern and $v(M^{\pm j}_p)$ are patterns with shifted labels as defined in \cite{vilenkin1992representation}. Both sums are running from $1$ to $[(p+1)/2]$. Moreover, if $p$ is even
\begin{align}
   A^j_p(M) &= \frac{1}{2} \left| \frac{\displaystyle \prod_{r=1}^{p/2-1}\left((l_{r,p-1}-1/2)^2-(l_{j,p}-1/2)^2\right)    \prod_{r=1}^{p/2}\left((l_{r,p+1}-1/2)^2-(l_{j,p}-1/2)^2\right)}{\displaystyle \prod_{r\neq j}(l^2_{r,p}-l^2_{j,p})(l^2_{r,p}-(l_{j,p}-1)^2)  } \right|^{\frac{1}{2}} \, ,\nonumber  \\
   l_{j,p} &= m_{j,p} + \frac{p}{2} -j \, ,
\end{align}
and if it is odd,
\begin{align}
   A^j_p(M) &=  \left| \frac{\displaystyle  \prod_{r=1}^{(p-1)/2}\left(l_{r,p-1}^2-l_{j,p}^2\right)    \prod_{r=1}^{(p+1)/2}\left(l_{r,p+1}^2-l_{j,p}^2\right)}{\displaystyle l^2_{j,p}(4l^2_{j,p}-1)\prod_{r\neq j}(l^2_{r,p}-l^2_{j,p})(l^2_{r,p}-(l_{j,p}-1)^2)  } \right|^{\frac{1}{2}} \, ,  \\
   l_{j,p} &= m_{j,p} + \frac{p+1}{2} -j \, .
\end{align}
The Gelfand-Tsetlin patterns and formulas for the action of generators on them provide a particular model for irreducible representations of $SO(n)$. In some applications, other models seem to be more useful, as was also the case in the present work. See \cite{Kologlu:2019mfz} for recent review of function space models.

\subsection{Tensor product decomposition of $SO(n)$ representations}
\label{subsec: Tensor product decomposition of $SO(n)$ representations}

All tensor products that we use as a part of procedure to generate matrix elements are of the type
\begin{equation}
 (k) \otimes \mu = \sum_i \nu_i \,,
\end{equation}
where $\mu$, $\nu_i$ are general irreducible representations of $SO(n)$ and $(k)$ an STT. These decompositions are governed by the following Pieri-type rule: remove $\frac{k+ |\mu|-|\nu|}{2}$ boxes from the Young diagram of $\mu$, and then add $\frac{k- |\mu|+|\nu|}{2}$ boxes to different columns to make $\nu$. Here, we are only interested in which representations appear in the decomposition, and not in their multiplicities.

\subsection{Representations of $O(n)$}

Here, we discuss the relation between irreducible representations of $SO(n)$ and $O(n)$. Our treatment follows the accounts of \cite{Brocker:1985:RCL,PROCTOR1994299}.
\smallskip

The analysis proceeds differently depending on whether $n$ is even or odd. Assume first that $n=2r+1$. In this case, the orthogonal group is of direct product form
\begin{equation}
    O(2r+1) \cong SO(2r+1) \times \mathbb{Z}_2\ .
\end{equation}
Therefore, irreducible representations of $O(2r+1)$ are tensor products of irreducible representations of the two factors, $(l_1,\dots,l_r)\otimes\pm$, conventionally denoted $(l_1,\dots,l_r)^\pm$.

Consider now the case of even $n$. Let $\pi=(l_1,\dots,l_r)$ be an irreducible representation of $SO(2r)$. We say that $\pi$ is of type I if $l_r=0$. Every representation of type I is the restriction of either of two non-isomorphic representations of $O(2r)$. These two representations will be denoted by $\pi^\pm=(l_1,\dots,l_r)^\pm$
\begin{equation}
    \text{Ind}_{SO(2r)}^{O(2r)} \pi = \pi^+ \oplus \pi^-, \qquad \text{Res}^{O(2r)}_{SO(2r)} \pi^+ = \text{Res}^{O(2r)}_{SO(2r)} \pi^- = \pi\ .
\end{equation}
Representations with $l_r\neq0$ are said to be of type II. Consider two such modules, $\pi^\pm = (l_1,\dots,l_{r-1},\pm l_r)$. Then they induce the same irreducible of $O(2r)$
\begin{equation}
    \pi = \text{Ind}_{SO(2r)}^{O(2r)} \pi^+ = \text{Ind}_{SO(2r)}^{O(2r)} \pi^-, \qquad \text{Res}^{O(2r)}_{SO(2r)}\pi = \pi^+ \oplus \pi^-\ .
\end{equation}
The above discussion can be extended to include for double covers of $SO(n)$ and $O(n)$ and spinorial representations. Restrictions between consecutive groups in all these cases are known, see the Proposition 10.1 of \cite{PROCTOR1994299}. 

\paragraph{Example} Representations of $O(4)$ are labelled either as $(J)^\pm$ or as $(J,q)$, with $J\geq q>0$ integers. Their restrictions to $SO(4)$ read
\begin{equation}
    \text{Res}^{O(4)}_{SO(4)}(J)^+ = \text{Res}^{O(4)}_{SO(4)}(J)^- = (J), \quad \text{Res}^{O(4)}_{SO(4)}(J,q) = (J,q) \oplus (J,-q)\ .
\end{equation}
Representations of $O(3)$ are labelled as $(l)^\pm$, where $l$ is a non-negative integer. Branching rules for STTs deoend on the sign label as
\begin{align}
    & \text{Res}^{O(4)}_{O(3)} (J)^+ = (J)^+ \oplus (J-1)^- \oplus \dots \oplus (0)^{(-)^J}\,,\\
    & \text{Res}^{O(4)}_{O(3)} (J)^- = (J)^- \oplus (J-1)^+ \oplus \dots \oplus (0)^{(-)^{(J+1)}}\ .
\end{align}

\section{Radial component map}
\label{Radial component map}

In this appendix, we define Harish-Chandra's radial component map and state Harish-Chandra's theorem related to it, \cite{HarishChandra,Warner2,Stokman:2020bjj,Reshetikhin:2020wep}. Let $G$ be a simple Lie group and $K\subset G$ a spherical subgroup. Consider a Cartan decomposition $G = K A_p K$ and let $h\in A_p$ be a generic element of $A_p$.\footnote{More precisely, $h$ should be in the regular part of $A_p$, see e.g. \cite{Reshetikhin:2020wep}. As it does not affect our discussion, we will not distinguish between $A_p$ and its regular part.} The universal enveloping algebra $U(\mathfrak{g}_c)$ admits a factorisation
\begin{equation}\label{U(g)-factorisations}
    U(\mathfrak{g}) \cong U(\mathfrak{a}_{p_c}) \otimes U(\mathfrak{k}_c) \otimes_{U(\mathfrak{m}_c)} U(\mathfrak{k}_c) \equiv U(\mathfrak{a}_{p_c}) \otimes \mathcal{K}_2\ .
\end{equation}
Here, $M$ denotes the centraliser of $A_p$ in $K$ and the subscript $c$ signifies complexification. We regard \eqref{U(g)-factorisations} as a family of vector space isomorphisms parametrised by elements $h$. Explicitly, the isomorphisms read
\begin{equation*}
    \Lambda_h : U(\mathfrak{a}_{p_c}) \otimes \mathcal{K}_2 \to U(\mathfrak{g}_c)\,, \quad \Lambda_h(H,x,y) = (h^{-1}x h)\ H\ y, \quad H\in U(\mathfrak{a}_{p_c}),\ x,y\in U(\mathfrak{k}_c)\ .
\end{equation*}
The preimage $\Lambda_h^{-1}(u)$ of some element $u$ from $U(\mathfrak{g}_c)$ is referred to as its radial decomposition with respect to $h$. By letting $h$ vary over $A_p$ and collecting all radial decompositions, one obtains a map
\begin{equation}\label{radial-component-map}
    \Pi: U(\mathfrak{g}_c) \to \text{Fun}(A_p) \otimes U(\mathfrak{a}_{p_c}) \otimes \mathcal{K}_2 \cong \mathcal{D}(A_p) \otimes \mathcal{K}_2\,,
\end{equation}
referred to as the Harish-Chandra's radial component map. In the second step, we used the fact that $A_p$ is abelian to identify $U(\mathfrak{a}_{p_c})$ with the algebra of differential operators on $A_p$ with constant coefficients. Consequently, the right-hand side is identified with (arbitrary) differential operators on $A_p$ with coefficients in $\mathcal{K}_2$. The significance of the map $\Pi$ for the theory of spherical functions lies in the following property. Let $u$ be an invariant differential operator on the group $G$ and $f$ a spherical function belonging to the space $\Gamma_{\rho,\sigma}$. We may regard $u$ as an element of the algebra $U(\mathfrak{g}_c)$. Then we have
\begin{equation}\label{Harish-Chandras-theorem}
    (u f)|_{A_p} = \left(\rho\otimes\sigma^\ast\right)\circ \Pi(u) \left(f|_{A_p}\right)\ .
\end{equation}
Let us elaborate on this formula. The element on the left-hand side is a vector-valued function on $A_p$, obtained by acting with $u$ on a spherical function and then restricting the result to the abelian subgroup $A_p$. The map on the right consists of two steps. Firstly, the element $u$ is radially decomposed in $U(\mathfrak{g}_c)$. Following the discussion below \eqref{radial-component-map}, we think of the result as a differential operator on $A_p$ with coefficients in $\mathcal{K}_2$. Secondly, abstract elements of the two copies of $U(\mathfrak{k}_c)$ that comprise $\mathcal{K}_2$ are evaluated in representations $\rho$ and $\sigma^\ast$. Therefore, $\Pi(u)$ may be regarded as the "universal restriction" of $u$ to $A_p$. If $u$ preserves a certain space of spherical functions $\Gamma_{\rho,\sigma}$, its reduction to this space is found by evaluating the universal restriction in the appropriate representations. However, \eqref{Harish-Chandras-theorem} is meaningful and holds true regardless of whether $u(\Gamma_{\rho,\sigma})\subset \Gamma_{\rho,\sigma}$ or not. The statement \eqref{Harish-Chandras-theorem} is Harish-Chandra's theorem.

\paragraph{Example} In the main text, we worked with the Gelfand pair $G = SO(d-1)$, $K = SO(d-2)$. Using the notation of Section \ref{Harmonic analysis}, the quadratic Casimir of $G$ can be written as
\begin{equation}\label{Casimir-1}
    C_2 = \frac12 L_{AB}L^{AB} = L_{12}L^{12} + L_{1i}L^{1i} + L_{2i}L^{2i} + \frac12 L_{ij}L^{ij}\ .
\end{equation}
Using the relations
\begin{equation}\label{radial-generators}
    L_{1i} = \cot\theta L_{2i} - \frac{L'_{2i}}{\sin\theta}, \quad [L_{2i},L'_{2i}] = -\sin\theta L_{12}\,,
\end{equation}
one finds the radial decomposition of the Casimir
\begin{equation}\label{radial-Casimir}
    C_2 = L_{12}^2 + (d-3)\cot\theta L_{12} + \frac{L'_{2i}L'_{2i} - 2\cos\theta L'_{2i}L_{2i} + L_{2i}L_{2i}}{\sin^2\theta} + \frac12 L^{ij}L_{ij}\ .
\end{equation}
The last relation and the theorem \eqref{Harish-Chandras-theorem} tell us that the reduction of the Laplacian to the space $\Gamma_{\rho,\sigma}$ is \eqref{main}. Meanwhile, the first equation in \eqref{radial-generators} forms the basis for the construction of external weight-shifting operators.

\section{External mixed symmetry tensors}
\label{External-mixed-symmetry-tensors}

In this appendix, we extend the solution theory to cases when representations $\rho$ and $\sigma$ are mixed symmetry tensors that have Young diagrams with two rows. This is the most general class of matrix elements that appear for $2\to2$ scattering of particles that are traceless symmetric tensors. We write $\rho = (l,\ell)$ and $\sigma = (l',\ell')$. The intermediate representation $\pi$ then has up to three non-vanishing GT labels $(J,q,s)$, which obey inequalities
\begin{equation}\label{labals-ordering}
    J \geq l,l' \geq q \geq \ell,\ell' \geq s\geq 0\ .
\end{equation}
The value of the quadratic Casimir in this representation is
\begin{equation}\label{Casimir-Jqs}
    C_2(J,q,s) = -J(J+d-3) - q(d+d-5) - s(s+d-7)\ . 
\end{equation}
Vectors $e^a$ and $e_\alpha$ can be labelled by Gelfand-Tsetlin patterns in the familiar way, which will now have pairs of potentially non-zero labels in each row - $m^{(1)}_{d-3}$, $m^{(2)}_{d-3}$, $m^{(1)}_{d-4}$, $m^{(2)}_{d-4}$, etc. By the same argument that we gave in Section \ref{Harmonic analysis}, in order to give a non-zero matrix element, $e^a$ and $e_\alpha$ have to have the same $SO(d-3)$ labels and furthermore the matrix element depends only on the labels $j_1 = m^{(1)}_{d-3}$ and $j_2 = m^{(2)}_{d-3}$
\begin{equation}\label{matrix-element-MST-MST}
    F^{d-1,J,q,s}_{l,\ell,l',\ell',j_1,j_2}(\theta) = \langle l,\ell;j_1,j_2;v|\pi_{J,q,s}(e^{\theta L_{12}})| l',\ell';j_1,j_2;v\rangle\ .
\end{equation}
Here, $v$ denotes the $SO(d-4)$-part of the pattern. Inequalities $l,l'\geq j_1\geq\ell,\ell'\geq j_2\geq0$ give the number of independent matrix elements
\begin{equation}\label{number-of-independent-me-MSTMST}
    N = \left(\text{min}(l,l') - \text{max}(\ell,\ell') + 1\right) \left(\text{min}(\ell,\ell')+1\right)\ .
\end{equation}
If $\ell=\ell'=0$, the above analysis reduces to the STT-STT case considered in the main text. The form of \eqref{number-of-independent-me-MSTMST} suggests that in the differential basis, the Laplacian assumes the form of a differential operator in one spacetime and two spin variables. If either of the representations $\rho$, $\sigma$ is a symmetric traceless tensor, the label $j_2$ disappears and one is left with a single spinning variable, a problem similar to the one with two STTs. These expectations are confirmed in the following.

\subsection{MST-STT system}
\label{subsec: MST-STT system}
In this subsection we treat the particular case when one of the two external representations reduces to an STT. First we construct the reduced Laplacian and then describe its solution theory.

\paragraph{Laplacian} Let us assume that $\sigma = (l')$ is an STT. We realise $\sigma$ by differential operators in the same way as was done in the main text. On the other hand, the representation $\rho=(l,\ell)$ is realised by
\begin{align}\label{MST-diff-rep-1}
    & \rho(L_{23}) = -i( x_a\partial_{x^a} - l), \quad \rho(L_{ab}) = x_a \partial_{x^b} - x_b\partial_{x^a} +z_a\partial_{z^b} - z_b\partial_{z^a}\,,\\
    & \rho(L_{2a}) = -\frac12 \Big((1-x^2)\partial_{x^a} + 2x_a (x^b\partial_{x^b} - l) + 2x^b (z_a\partial_{z^b} - z_b\partial_{z^a})\Big)\,,\\
    & \rho(L_{3a}) = \frac{i}{2}\Big((1+x^2)\partial_{x^a} - 2x_a(x^b\partial_{x^b}-l) - 2x^b (z_a\partial_{z^b} - z_b\partial_{z^a})\Big)\ .\label{MST-diff-rep-3}
\end{align}
These operators act on polynomials $F(x^a,z^a)$ which are of degree $\leq l$ in $x^a$ and degree $\ell$ in $z_a$. Furthermore, $F(x^a,z^a)$ are required to be harmonic in $z^a$, i.e. $\partial_{z^a}\partial_{z^a}F=0$. In practise, we first impose the condition $z^a\partial_{z^a}F = \ell F$ and then restrict to the lightcone $z^a z_a = 0$, as any homogeneous harmonic polynomial is uniquely determined by its values on the lightcone. For details, the reader is referred to \cite{Vilenkin:1993:RLG2,Bargmann_1977}. Similarly as for the STT-STT case, the $SO(d-3)$-invariance allows to reduce the Laplacian to an operator in variables $t$ and $y$. Since we go through all the steps in the more general MST-MST case below, here we only give the final result - when acting on functions of the form
\begin{equation}\label{prefactor-STT-MST}
    F = (1-X)^l (1-X')^{l'} y_2^\ell f(t,y)\,,
\end{equation}
the Laplacian is given by
\begin{equation}\label{MST-STT-Laplacian}
    \Delta^{(d)}_{l,\ell,l'} = \Delta^{(d)}_{l,l'} + \ell\ \frac{2(1+t^2-2ty)y\partial_y + (d+\ell-5)t^2 + 2(l+l'-\ell)t y +1}{1-t^2}\ .
\end{equation}
We remind the reader that $t=\cos \theta$ and the variable $y_2$ appearing in the prefactor in \eqref{prefactor-STT-MST} is defined below in \eqref{y2-definition}. 

\paragraph{Solution theory} Eigenfunctions $f^{J,q}_{l,\ell,l'}(t=\cos \theta,y)$ of \eqref{MST-STT-Laplacian} are constructed in a similar way as the ones for $\Delta^{(d)}_{l,l'}$ in Section \ref{solution theory STT-STT}. We can follow the same strategy, but we need to change appropriately the seed function and the external weight-shifting. The steps are summarised by the following diagram,
\vspace{6pt}

\begin{center}
\begin{tikzpicture}
    \node (f) at (3,0) {$f_{\ell,\ell,\ell}^{J,\ell}$};
    \node (g) at (6,0) {$f_{l,\ell,\ell}^{J,\ell}$};
    \node (h) at (9,0) {$f_{l,\ell,l'}^{J,\ell}$};
    \node (i) at (12,0) {$f_{l,\ell,l'}^{J,q} \, .$};
    
    \draw[->] (f) -- node[above] {$(q)^{l-\ell}$} (g);
    \draw[->] (g) -- node[above] {$(\bar{q})^{l'-\ell}$} (h);
    \draw[->] (h) -- node[above] {internal} (i);
    \draw[->] (h) -- node[below] {w.s.} (i);
\end{tikzpicture}
\end{center}

More specifically:
\begin{itemize}  
\item To obtain the seed function, we solve the differential equation in the boundary case $\Delta_{\ell,\ell,\ell}^{(d)}$ with the eigenvalue $(J,\ell)$. This function is independent of $y$,
\begin{equation}\label{seed-function-MST-STT}
    f^{J,\ell}_{\ell,\ell,\ell}(t) = (1-t^2)^{\ell/2} C_{J-\ell}^{\left(\frac{d-3+2\ell}{2}\right)}(t)\ .
\end{equation}

\item Then, as before, we need to act with external weight-shifting operators, which depend additionally on $\ell$,
\begin{equation}\label{weight-shifitng-MST-STT}
    q_{l,\ell,l'} = \partial_\theta - l \cot\theta + \frac{(y^2-1)\partial_y + (\ell-l')y}{\sin\theta}, \quad \bar q_{l,\ell,l'} = \partial_\theta - l' \cot\theta + \frac{(y^2-1)\partial_y + (l-\ell)y}{\sin\theta}\ .
\end{equation}
The exchange relations \eqref{shift-equations} hold for operators $\Delta^{(d)}_{l,\ell,l'}$, $q_{l,\ell,l'}$, $q_{l,\ell,l'}$ as well - one should simply add the same label $\ell$ to all operators appearing in \eqref{shift-equations} (this quantum number is not shifted by commuting $q,\bar q$ past $\Delta$). After the action of external weight-shifting operators, we get solution for $\Delta_{l,\ell,l'}^{(d)}$ with eigenvalues $(J,\ell)$, $f^{J,\ell}_{l,\ell,l'}(t)$.

\item The last step involves the action of internal weight-shifting operators (see Subsection \ref{subsubsec: Internal weight-shifting}), which is the same as the STT-STT case with $\Delta_{l,\ell,l'}^{(d)}$ instead of $\Delta_{l,l'}^{(d)}$. This will get us the general solution for $\Delta_{l,\ell,l'}^{(d)}$ with eigenvalues $(J,q)$, $f^{J,q}_{l,\ell,l'}(t)$. In the rest of the example, we will not care about normalisation, but only about the dependence on $(t,y)$. Hence, we have a proportionality sign '$\propto$', instead of an equality. We can do this without lost of generality, because we can deal with the normalisation of the matrix elements only at the very end.
\end{itemize}

\paragraph{Example}\label{eg: MST-STT} Let us now compute a particular eigenfunction of the Laplacian as an illustration of the above method. Suppose we want to know $f^{3,2}_{3,1,2}(t,y)$ \footnote{In principle we could decide not to fix $J$.} (notice that the quantum numbers should satisfy \eqref{labals-ordering} to obtain a non-zero value).
\begin{itemize}
    \item Let us start from the seed function \eqref{seed-function-MST-STT}, $f^{3,1}_{1,1,1}(t)\propto \sqrt{1-t^2} \left( (d+1)t^2 -1 \right) $.
    \item We act twice with $q$ and once with $\bar{q}$, \eqref{weight-shifitng-MST-STT}, in order to get to $l=3$ and $l'=2$,
    \begin{equation}
       f^{3,1}_{3,1,2}(t,y) = q_{2,1,2} \, q_{1,1,2} \, \bar q_{1,1,1} \, f^{3,1}_{1,1,1}(t) \propto (t^2-1)(t-y)  \, .
    \end{equation}
    \item The last step is to do the internal weight-shifting in order to obtain $q=2$. We need to multiply the previous result with the zonal spherical function $f^1_{0,0}(t)$, \eqref{zonal-spherical}, and compute the tensor decomposition $(3,1) \otimes (1) = (3) \oplus (2,1) \oplus (4,1) \oplus (3,2) $. Then, we project out the 'unwanted' representations to obtain the non-normalised matrix element,
    \begin{small}
        \begin{align*}
     f^{3,2}_{3,1,2}(t,y) &= \left(\Delta_{3,1,2}^{(d)} - C_2(4,1)\right) \left((\Delta_{3,1,2}^{(d)} - C_2(2,1)\right) \left(\Delta_{3,1,2}^{(d)} - C_2(3,0)\right)( f^{1}_{0,0}(t) \, f^{3,1}_{3,1,2}(t,y) )\\
     &\propto (t^2-1)(t y-1) \, .
    \end{align*}
    \end{small}
    
\end{itemize}

\subsection{MST-MST system}
\label{subsec: MST-MST system}
Now we deal with the general case where both external representations are mixed symmetry tensors that have Young diagrams with two rows. This is quite different with respect to previous cases, since the Laplacian depends on an extra spin variable. Hence, both the reduction and the solution theory for this system will be more involved.

\paragraph{Laplacian} To derive the reduced Laplacian when $\rho= (l,\ell)$ and $\sigma= (l',\ell')$, we use \eqref{main}, substitute \eqref{MST-diff-rep-1}-\eqref{MST-diff-rep-3} for the representation $\rho$, and similarly for $\sigma$. Then the condition \eqref{SO(d-4)-invariance} requires functions to depend only on scalar products
\begin{align*}
   & X = x^2,\quad X' = x'^2,\quad Z = z^2, \quad Z' = z'^2, \quad Y = x\cdot z\,,\\
   & Y' = x'\cdot z', \quad U = x\cdot z',\quad U' = x'\cdot z, \quad W = x\cdot x', \quad T = z\cdot z'\ .
\end{align*}
Imposing the full $M$-invariance \eqref{m-invariance} further restricts to functions of the form
\begin{equation*}
    F = (1-X)^l (1-X')^{l'} F\left(y,y_1,y_2,y'_2\right)\ .
\end{equation*}
Here, $y$ is the same variable as introduced in Section \ref{Harmonic analysis}, the $y_{1,2}$ read 
\begin{align}\label{y1-definition}
    & y_1 = - T - \frac{2(U'-Y)(U-Y')}{X+X'-2W}\equiv \frac{\Theta_1}{X+X'-2W}\,,\\
    & y_2 = \frac{2WY-X'Y-XU'-Y+U'}{(1-X)(1-X')}\equiv\frac{\Theta_2}{(1-X)(1-X')}\,,\label{y2-definition}
\end{align}
and $y'_2$ is obtained from $y_2$ by exchanging the roles of primed and unprimed coordinates (with $W'=W$ and $T'=T$). Sometimes, instead of $y$ it will be more convenient to use the variable $y_0$ related to it by $y=2y_0+1$. Finally, we impose conditions $z_a\partial_{z_a} F = \ell F$, $z'_a\partial_{z'_a} F = \ell' F$ and restrict to lightcones $z^a z_a = z'^a z'_a = 0$ - recall that these conditions are a part of the function space description for the MST representation. In the end, one is left with functions of the form\footnote{Without loss of generality, in the following formulas we assume $\ell\leq\ell'$.}
\begin{equation}\label{prefactor-MST-MST}
    F = (1-X)^l (1-X')^{l'}\, y_2^\ell\, y'^{\ell'}_2\, (x-y)^{-\ell}\, f\left(x,y\right), \quad x = 1 + \frac{2 y_0 y_1^2}{y_0 y_1 + 2 y_2 y'_2}\ .
\end{equation}
The function $F$ must be a polynomial of polarisation vectors $z_a$ and $z'_a$ - the space of allowed functions is thus finite-dimensional and spanned by monomials $x^m y^n$ with
\begin{equation}
    m\leq \text{min}(\ell,\ell')\, ,\quad  \quad n\leq \text{min}(l,l')-\text{max}(\ell,\ell')\ .
\end{equation}
In the {\it Mathematica} notebook, we give the Laplacian $\Delta^{(d)}_{l,\ell,l',\ell'}$. In writing it, it is useful to introduce
\begin{equation}\label{L-shadow-xy}
    \mathcal{D}^{(d)}_{x,y} = \mathcal{D}^{(d-2)}_x + \mathcal{D}^{(d+2\ell')}_y + 2(x^2-1)\partial_x\partial_y - 2\ell x\partial_y + \ell(\ell+d-5)\ .
\end{equation}
This operator generalises the Gegenbauer differential operator \eqref{Gegenbauer-operator} to two variables and will play an important role below. In terms of $\mathcal{D}^{(d)}_{x,y}$, the Laplacian assumes the form
\begin{small}
    \begin{align}
    \Delta^{(d)}_{l,\ell,l',\ell'} & = \partial_\theta^2 + (d-3)\cot\theta \partial_\theta - \mathcal{D}^{(d)}_{x,y}\nonumber\\
    & + \frac{2\mathcal{D}^{(d)}_{x,y}-l(l+d-4)-l'(l'+d-4)-\ell(\ell+d-6)-\ell'(\ell'+d-6)}{\sin^2\theta}\label{MST-MST-Laplacian}\\
    & -2\cos\theta \frac{x \mathcal{D}^{(d-2)}_x + y \mathcal{D}^{(d+2\ell)}_y - (l+\ell+l'-\ell'+d-6)(x^2-1)\partial_x - (l+l'+d-5)(y^2-1)\partial_y}{\sin^2\theta}\nonumber\\
    & -2\cos\theta \frac{2(x^2-1)y \partial_x\partial_y - 2\ell x y \partial_y +\ell(l+l'-\ell'+1)x + (l-\ell')(l'-\ell')y}{\sin^2\theta}\ .\nonumber
\end{align}
\end{small}

It preserves the space of functions $f(t,x,y)$ polynomial in $x$ and $y$ as above. Moreover, notice that $\Delta^{(d)}_{l,\ell,l',\ell'}$ is consistent with the previous MST-STT result when $\ell'=0$, acting on functions $f(t,y)$,
\begin{equation}\label{boundary-case-1}
    \Delta_{l,\ell,l',0} f(\theta,y) = \Delta^{MST-STT}_{l,\ell,l'} f(\theta,y)\ .
\end{equation}

\paragraph{Solution theory} In this case we are dealing with a differential equation in three variables $(t,x,y)$, which makes the strategy more difficult. Nevertheless, we want to proceed in the usual way: we solve for a boundary case to obtain a seed function, and then use weight-shifting operators. We summarise the steps by the following diagram,
\vspace{6pt}

\begin{center}
\begin{tikzpicture}
    \node (e) at (0,0) {$f_{l,0,\ell',\ell'}^{J,\ell',0}$};
    \node (f) at (3,0) {$f_{l,\ell,\ell',\ell'}^{J,\ell',0}$};
    \node (g) at (6,0) {$f_{l,\ell,l',\ell'}^{J,\ell',0}$};
    \node (h) at (9,0) {$f_{l,\ell,l',\ell'}^{J,q,s} \, .$};
    
    \draw[->] (e) -- node[above] {$(q_{\scriptscriptstyle MST-STT})^{\ell}$} (f);
    \draw[->] (f) -- node[above] {$(\bar{q})^{l'-\ell'}$} (g);
    \draw[->] (g) -- node[above] {internal} (h);
    \draw[->] (g) -- node[below] {w.s.} (h);
\end{tikzpicture}
\end{center}

More precisely:
\begin{itemize}

    \item The first problem is that we cannot go the same way as before in order to obtain the seed function. Now we are not able to identify any particular case with no dependence on both $x$ and $y$. The way to proceed is to notice that in the boundary case $l'=\ell'$, the operator \eqref{MST-MST-Laplacian} is well-defined on functions $f(t,x)$ and assumes the form
\begin{equation}\label{boundary-case-2}
    \Delta_{l,\ell,\ell',\ell'} f(\theta,x) = \left(\Delta^{MST-STT}_{l,-1,\ell-1} - (\ell'+1)(\ell'+d-6)\right) f(\theta,x)\ .
\end{equation}
    This is the crucial new ingredient that allows for a complete solution theory. 
    
    \item The next step is to solve the right hand side of \eqref{boundary-case-2}. The strategy is to look for a boundary solution for the MST-STT system, which is similar to the previous case in Section \ref{subsec: MST-STT system}, but with a different constant term. This suggests us to use as the seed function the boundary case where $\ell -1=-1$, $q=\ell'$ and $s=0$, whose solution is
    \begin{equation}\label{seed-function-MST-MST}
    f^{J,\ell',0}_{l,0,\ell',\ell'}(\cos \theta) = (1-\cos^2\theta)^{l/2} C_{J-l}^{\left(\frac{d-3+2l}{2}\right)}(\cos \theta)\ .
    \end{equation}
    At this point, we can raise the value $\ell$ by the action of MST-STT external weight-shifting operators \eqref{weight-shifitng-MST-STT}\footnote{Differently from the MST-STT case, these weight-shifting operators depends on $(t,x)$ and not $(t,y)$.}. This is easy to understand if you remember that this is an eigenfunction for the right hand side of \eqref{boundary-case-2}, which has the appropriate commutation relations with \eqref{weight-shifitng-MST-STT}. Hence, at this point, we have a solution for $\Delta_{l,\ell,\ell',\ell'}$ with eigenvalues $(J,\ell',0)$, $f^{J,\ell',0}_{l,\ell,\ell',\ell'}(\cos \theta,y)$.

    \item The action of external and internal weight-shifting operators is as in Section \ref{subsec:Weight-shifting operators}, with the difference that now they depend on $x$, too,
\begin{align}
    & q_{l,\ell,l',\ell'} = \partial_\theta - l \cot\theta + \frac{(x^2-1)\partial_x + (y^2-1)\partial_y - \ell x + (\ell'-l')y}{\sin\theta}\,,\label{WS-MST-MST-1}\\
    & \bar q_{l,\ell,l',\ell'} = \partial_\theta - l' \cot\theta + \frac{(x^2-1)\partial_x + (y^2-1)\partial_y - \ell x + (\ell'-l)y}{\sin\theta}\label{WS-MST-MST-2}\ .
\end{align}
    Laplacians and weight shifting operators satisfy the same exchange relations as before - the labels $\ell$ and $\ell'$ are just added to both sides of \eqref{shift-equations} and 'go along for the ride'
\begin{equation}\label{exchange-relations}
    q_{l,\ell,l',\ell'} \Delta^{(d)}_{l,\ell,l',\ell'} = \Delta^{(d)}_{l+1,\ell,l',\ell'} q_{l,\ell,l',\ell'}, \qquad \bar q_{l,\ell,l',\ell'} \Delta^{(d)}_{l,\ell,l',\ell'} = \Delta^{(d)}_{l,\ell,l'+1,\ell'} \bar q_{l,\ell,l',\ell'}\ .
\end{equation}
    In practice, we use the two types weight-shifting in turn, first for $f^{J,\ell',0}_{l,\ell,\ell',\ell'}(t,y)\rightarrow f^{J,\ell',0}_{l,\ell,l',\ell'}(t,y,x)$, and then to get to the general solution $f^{J,q,s}_{l,\ell,l',\ell'}(t,y,x)$.
    
\end{itemize}

\paragraph{Example} Let us now do an example of the previous method computing $f^{4,2,1}_{4,1,3,2}(t=\cos \theta,x,y)$\footnote{As in previous Example \ref{eg: MST-STT}, here we will only deal with non-normalised matrix elements. Hence, we drop eventual numerical factors.}. 
\begin{itemize}
    \item The seed function \eqref{seed-function-MST-MST} for our case is $f^{4,2,0}_{4,0,2,2}(t)\propto(t^2-1)^2$. Then, we act with MST-STT external weight-shifting to obtain $f^{4,2,0}_{4,1,2,2}(t,x)\propto(1-t^2)^\frac{3}{2}(t-x)$.
    \item At this point, we act with MST-MST external weight-shifting operators \eqref{WS-MST-MST-2} to get to $f^{4,2,0}_{4,1,3,2}(t,x,y)\propto(t^2-1)(t-x)(t-y)$. Finally, we use exactly as in previous Example \ref{eg: MST-STT} internal weight-shifting to obtain the non-normalised $f^{4,2,1}_{4,1,3,2}(t,x,y)\propto(t^2-1)(tx-1)(t-y)$.
\end{itemize}

\section{From polynomials to the Gelfand-Tsetlin basis}
\label{From polynomials to the Gelfand-Tsetlin basis}

In this appendix, we show how to change between differential basis used for most of our computations and the more usual Gelfand-Tsetlin basis. Assume the dimension $d$, as well as the external and internal labels have been fixed and that we have computed the matrix element in the form of the function $f(\theta,x,y)$ as explained in the previous sections. Upon reinstating the prefactor as in \eqref{prefactor-MST-MST}, we obtain the function
\begin{small}
    \begin{equation*}
    F(\theta,x,z,x',z') = \pi^a{}_\alpha(e^{\theta L_{12}}) e_a \otimes e^\alpha = \sum_{j_1,j_2;v} \langle j_1,j_2,v |\pi(e^{\theta L_{12}})| j_1,j_2;v\rangle g_{j_1,j_2,v}(x,z) g'_{j_1,j_2,v}(x',z')\ .
\end{equation*}
\end{small}

We have suppressed parts of GT patterns that are fixed by the choice of external and internal quantum numbers, $(l,\ell,l',\ell')$ and $(J,q,s)$. Functions $g_{j_1,j_2,v}(x,z)$ are the vectors $|j_1,j_2;v\rangle$ written in the function-space realisation of the representation $(l,\ell)$ that we have given above. A similar comment applies to $g'_{j_1,j_2,v}(x',z')$. Since the matrix elements are independent of $v$, we can split the above sum to obtain 
\begin{equation}
    F(\theta,x,z,x',z') = \sum_{j_1,j_2} F_{j_1,j_2}(\theta)\sum_v g_{j_1,j_2,v}(x,z) g'_{j_1,j_2,v}(x',z')\ .
\end{equation}
The matrix elements we are interested in are functions $F_{j_1,j_2}(\theta)$. The last sum runs over all states $v$ in the representation $(j_1,j_2)$ of $SO(d-3)$. It produces an $SO(d-3)$-invariant vector in the tensor product $\rho\otimes\sigma^\ast$. This vector is of the functional form \eqref{prefactor-MST-MST}, but of course independent of $\theta$. From these remarks, the solution $f(\theta,x,y)$ obtained through weight-shifting decomposes as
\begin{equation}
    f(\theta,x,y) = \sum_{j_1,j_2} F_{j_1,j_2}(\theta) P_{j_1,j_2}(x,y)\,,
\end{equation}
where $P_{j_1,j_2}(x,y)$ are appropriate polynomials. The latter form a basis for the space of polynomials in $x$ and $y$, with with degrees up to $\text{min}(l,l')-\text{max}(\ell,\ell')$ and $\text{max}(\ell,\ell')$, respectively. They are eigenfunctions of the quadratic Casimir of $SO(d-3)$ in the representation $\rho$. This operator was in fact computed above - it is (minus) $\mathcal{D}^{(d)}_{x,y}$ of \eqref{L-shadow-xy}. Thus, $P_{j_1,j_2}$ are determined by solving
\begin{equation}\label{GT-polynomials}
    \mathcal{D}^{(d)}_{x,y} P_{j_1,j_2}(x,y) = -C_2^{SO(d-3)}(j_1,j_2) P_{j_1,j_2}(x,y)\ .
\end{equation}
In the case of symmetric traceless tensors the equation \eqref{GT-polynomials} specialises to the Gegenbauer differential equation
\begin{equation}\label{GT-polys-STT-STT}
    \mathcal{D}^{(d)}_y P_j(y) = j(j+d-5) P_j (y) \quad\implies\quad P_j(y) = C^{\left(\frac{d-5}{2}\right)}_j(y)\ .
\end{equation}
In the mixed MST-STT case, we have $j_2 = 0$ and independent matrix elements are written as $F_j(\theta)$. Polynomials $P_j$ are again solutions to the Gegenbauer equation, now with shifted parameters
\begin{small}
    \begin{equation}\label{GT-polys-MST-STT}
    \mathcal{D}^{(d+2\ell)}_y P_j(y) = \left( j(j+d-5) - \ell(\ell+d-5) \right) P_j(y) \quad\implies\quad P_j(y) = C^{\left(\frac{d+2\ell-5}{2}\right)}_{j-\ell}(y)\ .
\end{equation}
\end{small}

In the general MST-MST, we do not provide a closed-form expressions for polynomials $P_{j_1,j_2}(x,y)$, but rather compute them on case-by-case basis. For purposes of applications, this is a simple matter, as the dimension \eqref{number-of-independent-me-MSTMST} of the space of these polynomials is quite small.

\subsection{Normalisation}

The method described above computes functions $F^{J,q,s}_{l,\ell,l',\ell',j_1,j_2}(\theta)$ up to normalisation. In this section, we explain how the functions are normalised. With this in mind, let
\begin{equation}
    f^{J,q,s}_{l,\ell,l',\ell',j_1,j_2}(\theta) = c^{J,q,s}_{l,\ell,l',\ell',j_1,j_2} F^{J,q,s}_{l,\ell,l',\ell',j_1,j_2}(\theta)\,,
\end{equation}
be a function function proportional to the matrix element \eqref{matrix-element-MST-MST}, with some arbitrary proportionality coefficient $c$ that depends on all internal and external quantum numbers, as well as the dimension $d$. Our idea is very simple - we expand the function $f$ in $\theta$ and pick the first non-vanishing term. The corresponding term of $F$ is calculated from the Gelfand-Tsetlin formulas, \cite{vilenkin1992representation}. Indeed, upon expanding \eqref{matrix-element-MST-MST}, we obtain
\begin{small}
    \begin{align}\label{normalisation-expansion}
    F^{J,q,s}_{l,\ell,l',\ell',j_1,j_2}(\theta) & = \langle l,\ell;j_1,j_2;v|1 + \theta L_{12} + \dots| l',\ell';j_1,j_2;v\rangle\\
    & = \delta_{ll'} \delta_{\ell\ell'} + \theta \langle l,\ell;j_1,j_2;v| L_{12} | l',\ell';j_1,j_2;v\rangle + \frac{\theta^2}{2}\langle l,\ell;j_1,j_2;v| L^2_{12} | l',\ell';j_1,j_2;v\rangle + \dots\ .\nonumber
\end{align}
\end{small}

If the left and right representations are the same, $(l,\ell) = (l',\ell')$, the expansion starts at the zeroth order and the coefficient $c$ is computed as $c = f(\theta=0)$. In general, the power of $\theta$ at which the expansion starts is the "distance" between representations $(l,\ell)$ and $(l',\ell')$, namely
\begin{equation}
    d_{(l,\ell),(l',\ell')} = |l-l'| + |\ell-\ell'|\ .
\end{equation}
This follows inductively from the fact that $L_{12}(l,\ell)$ has non-zero overlaps with only the four vectors $(l\pm1,\ell),(l,\ell\pm1)$, \eqref{Action-infinitesimal-generators}. If the distance is one, the linear term in \eqref{normalisation-expansion} is directly computed by the GT formula. For distances $d_{(l,\ell),(l',\ell')}>1$, one sums over all "admissible paths" from $(l,\ell)$ to $(l',\ell')$. We give one example. Let $l'=l+1$ and $\ell' = \ell+1$. Then, suppressing indices $j_1,j_2,v$ on the right,
\begin{small}
    \begin{align*}
    & F^{J,q,s}_{l,\ell,l+1,\ell+1,j_1,j_2}(\theta) = \frac{\theta^2}{2}\sum_{\text{int}}\langle l,\ell| L_{12} |\text{int}\rangle\langle\text{int}| L_{12}| l+1,\ell+1\rangle + \dots\\
    & = \frac{\theta^2}{2}\Big(\langle l,\ell| L_{12} |l,\ell+1\rangle\langle l,\ell+1| L_{12}| l+1,\ell+1\rangle +\langle l,\ell| L_{12} |l+1,\ell\rangle\langle l+1,\ell| L_{12}| l+1,\ell+1\rangle \Big) + \dots\ .
\end{align*}
\end{small}

Clearly, no other intermediate states int contribute to the sum. All expressions in the last line are determined by GT formulas \eqref{Action-infinitesimal-generators}. For higher values of $d_{(l,\ell),(l',\ell')}$ one continues in the obvious way with the help of some elementary combinatorics.

\section{Practical implementation of the algorithm}
\label{Practical implementation of the algorithm}

We have prepared a file \href{https://gitlab.com/russofrancesco1995/partial\_waves}{ gitlab.com/russofrancesco1995/partial\_waves} where all the constructions discussed above are implemented. In this appendix we explain in detail how to use the most important functions. For readers interested only in using the final results, we point to Section \ref{subsec: Functions to generate matrix elements}. The others subsections describe the functions for each different case.

\subsection{Functions to generate matrix elements}
\label{subsec: Functions to generate matrix elements}
In practice, we can divide the process to obtain partial waves into two steps. At first, we need to understand which external irreducible representations appear in the scattering. Then, compute the matrix elements with those representations. The first step is easily done, since it consists in a tensor decomposition\footnote{See Sections \ref{subsec:Irreducible content of two-particle states} and \ref{subsec: Tensor product decomposition of $SO(n)$ representations} for more details.}, while we give here the solution on how to obtain the matrix elements,
\begin{small}
    \begin{align}\label{GenerateFunctions-nb}
    \text{STT-STT case:}  \qquad &F^{J,q}_{l,l'}(\theta) = \texttt{Generate\$STTSTT\$GTBasis[d,l,l',J,q][$\theta$]}\,,\\
    \text{MST-STT case:}  \qquad &F^{J,q}_{l,\ell,l'}(\theta) = \texttt{Generate\$MSTSTT\$GTBasis[d,l,l',$\ell$,J,q][$\theta$]}\,,\\
    \text{MST-MST case:}  \qquad &F^{J,q,s}_{l,\ell,l',\ell'}(\theta) = \texttt{Generate\$MSTMST\$GTBasis[d,l,l',$\ell$,$\ell'$,J,q,s][$\theta$]}\ .
\end{align}
\end{small}

To run this functions, one needs to fix the dimension $d$ and all quantum numbers, while $\theta$ is the scattering angle and can be left as a variable. These functions give the normalised matrix elements in the Gelfand-Tsetlin basis, as described in Appendix \ref{From polynomials to the Gelfand-Tsetlin basis}. Notice that in the first two cases the result is a vector, while in the MST-MST one we have a matrix. This is explained by the fact that in the last case the external states have an extra label, which is 0 when one of the state is a STT. Moreover, the function for generating matrix elements for the MST-MST case in \eqref{GenerateFunctions-nb}, without lost of generality is well-defined only if $\ell'\geq \ell$. We can always recover the opposite scenario by doing a complex conjugation of the result.

\subsection{STT-STT functions}
In the case of external STT-STT representations, the solution theory for matrix elements is different in the two cases, $d=5$ and $d>5 \,$\footnote{See Section \ref{solution theory STT-STT} and Section \ref{Exceptions in five dimensions} for details on the theory}. However, this difference is automatically implemented in the functions in the choice of the dimension\footnote{This is not the case when the function is defined in only one of the cases, e.g. the function for the second Casimir is defined only in $d=5$.}. Hence, even if they do different things, the user will be able to treat in mostly the same way the two cases. 

\paragraph{Definitions}The first function is the reduced quadratic Casimir \eqref{final} acting on a general function $f(t=\cos \theta,y)$,
\begin{equation*}\label{laplacianSTTSTT-nb}
    \Delta^{(d)}_{l,l'}f(t,y)= \texttt{ReducedLaplacian\$STTSTT[d,l,l'][t,y][f]}\ .
\end{equation*}
In $d=5$, we need to compute the second Casimir \eqref{eq: Casimirs in 5d} for the action of the internal weight-shifting operators. Hence, in a similar way we have that the reduced second Casimir minus the eigenvalue is $\texttt{ReducedSecondCasimirEigenvalue\$STTSTT\$5d[l,l',J,q][t,y][f]}$.
The external weight-shifting operators \eqref{weight-shifting} are defined as 
\begin{align*}
    q_{l,l'}f(t,y) &= \texttt{qExternalWeightShifting\$STTSTT[l,l'][t,y][f]}\,,\\
    \bar{q}_{l,l'}f(t,y) &= \texttt{qbExternalWeightShifting\$STTSTT[l,l'][t,y][f]}\ .
\end{align*}

\paragraph{Solution theory in differential basis}
In this paragraph, we describe the functions to find non-normalised matrix elements in the differential basis. This means that we find a solution to the Laplacian, and one can always check to have the correct one by acting with \eqref{laplacianSTTSTT-nb} minus the correct eigenvalue. We have defined the action of the external and internal weight-shifting operators (Section \ref{subsec:Weight-shifting operators}) as
\begin{align*}
    f^{J,0}_{l,l'}(t,y) = & \texttt{ActionExternalWeightShifting\$STTSTT[d,l,l',J][t,y]}\,,\\
    f^{J,q}_{l,l'}(t,y) = & \texttt{ActionInternalWeightShifting\$STTSTT[d,l,l',J,q][t,y][$f^{J,0}_{l,l'}(t,y)$]}\,,
\end{align*}
and the two actions together as $f^{J,q}_{l,l'}(t,y)=\texttt{Generate\$STTSTT\$yBasis[d,l,l',J,q][t,y]}$.

\paragraph{Map to Gelfand-Tsetlin basis and normalisation}
We want to take the previous result and express it in the Gelfand-Tsetlin basis, Section \ref{subsec: Matrix elements in the Gelfand-Tsetlin basis and checks}. The map differential basis $\rightarrow$ Gelfand-Tsetlin basis is implemented as $\texttt{toGTbasis\$STTSTT[d][y][f(t,y)]}$, where $f(t,y)$ is a polynomial in $y$, with functions of $t$ as coefficients. Now, we are able to normalise our matrix elements with $\texttt{normaliseSTTSTTinGT[d,l,l',J,q][$\theta$][f]}$, and we put these two steps together in the function previously introduced in \eqref{GenerateFunctions-nb}. Notice that the result is a vector of length
\begin{itemize}
    \item $\text{min}(l,l')+1$, if $d>5\,$,
    \item $2 \, \text{min}(l,l')+1$, if $d=5\ $.
\end{itemize}

\subsection{MST-STT functions}
The functions which we will describe in this subsection are defined only for $d \geq 8$, since the exceptional cases have not been implemented yet. Recall that in $d=5$ we have no MST-representations, while in $d=6,7$ one should make slight modifications in our computations, analogous to what was done in Section \ref{Exceptions in five dimensions} for $d=5$ STT-STT matrix elements. We remind to Appendix \ref{subsec: MST-STT system} for details on the theory.

\paragraph{Definitions} The names of the functions are the same as in the STT-STT case, with the replacement of STT-STT by MST-STT. The other difference with the previous case is that we have an extra index, since MST-representations needs an extra one to be well-defined.
\begin{itemize}
    \item The reduced quadratic Casimir (Laplacian) \eqref{MST-STT-Laplacian} is 
    $$\Delta^{(d)}_{l,\ell,l'}f(t,y)= \texttt{ReducedLaplacian\$MSTSTT[d,l,l',$\ell$][t,y][f]}\ .$$
    \item The external weight-shifting operators \eqref{weight-shifitng-MST-STT} are
    \begin{align*}
    q_{l,\ell,l'}f(t,y) &= \texttt{qExternalWeightShifting\$MSTSTT[l,l',$\ell$][t,y][f]}\,,\\
    \bar{q}_{l,\ell,l'}f(t,y) &= \texttt{qbExternalWeightShifting\$MSTSTT[l,l',$\ell$][t,y][f]}\ .
    \end{align*}
    \item The zonal spherical function is $f^{J,0}_{0,0}(t)=\texttt{zonalSphericalFunction[d,J][t]}$, and the seed solution \eqref{seed-function-MST-STT}, $f^{J,\ell}_{\ell,\ell,\ell}(t)=\texttt{boundarySolution\$MSTSTT[d,J,$\ell$][t]}$.
\end{itemize}

\paragraph{Solution theory}
The functions are the same as in the STT-STT case, for completeness we list them here:
\begin{itemize}
    \item Action of external and internal weight-shifting operator in differential basis, respectively,
\begin{small}
    \begin{align*}
    f^{J,\ell}_{l,\ell,l'}(t,y) = & \texttt{ActionExternalWeightShifting\$MSTSTT[d,l,l',$\ell$,J][t,y]}\,,\\
    f^{J,q}_{l,\ell,l'}(t,y) = & \texttt{ActionInternalWeightShifting\$MSTSTT[d,l,l',$\ell$,J,q][t,y][$f^{J,\ell}_{l,l'}(t,y)$]}\ .
\end{align*}
\end{small}
        
    \item From polynomial in $y$, to Gelfand-Tsetlin basis, $\texttt{toGTbasis\$MSTSTT[d,$\ell$][y][f]}$.
    \item Normalisation of matrix elements, $\texttt{normaliseMSTSTTinGT[d,l,l',$\ell$,J,q][$\theta$][f]}$.
\end{itemize}
Notice that the result is a vector with length $\text{min}(l,l')-\ell +1$.

\subsection{MST-MST functions}
As for the previous case, the functions for the MST-MST case are well-defined only for $d\geq 8$. Here, we only list them, since the strategy is the same and should be clear from the previous subsections. Let us stress again the fact that some functions here are well-defined only if $\ell' \geq \ell$. The opposite case can always be obtained by a complex conjugation\footnote{More detail on the theory can be found in Appendix \ref{subsec: MST-MST system}.}.

\begin{itemize}
    \item The reduced quadratic Casimir (Laplacian) \eqref{MST-MST-Laplacian} is 
    $$\Delta^{(d)}_{l,\ell,l',\ell'}f(t,y,x)= \texttt{ReducedLaplacian\$MSTMST[d,l,l',$\ell$,$\ell'$][t,y,x][f]}\ .$$
    \item The external weight-shifting operators \eqref{WS-MST-MST-1} are
    \begin{align*}
    q_{l,\ell,l',\ell'}f(t,y,x) &= \texttt{qExternalWeightShifting\$MSTMST[l,l',$\ell$,$\ell'$][t,y,x][f]}\,,\\
    \bar{q}_{l,\ell,l',\ell'}f(t,y,x) &= \texttt{qbExternalWeightShifting\$MSTMST[l,l',$\ell$,$\ell'$][t,y,x][f]} \, .
    \end{align*}
    \item The zonal spherical function as before is\\ $f^{J,0,0}_{0,0,0,0}(t)=\texttt{zonalSphericalFunction[d,J][t]}$, and the seed solution \eqref{seed-function-MST-MST},\\ $\texttt{boundarySolution\$MSTMST[d,J,l][t]}$.

    \item Action of external and internal weight-shifting operator in differential basis, respectively,
    \begin{small}
            \begin{align*}
    &f^{J,\ell',0}_{l,\ell,l',\ell'}(t,y,x) =  \texttt{ActionExternalWeightShifting\$MSTMST[d,l,l',$\ell$,$\ell'$,J][t,y,x]}\,,\\
    &f^{J,q,s}_{l,\ell,l',\ell'}(t,y,x) = \\ &\texttt{ActionInternalWeightShifting\$MSTMST[d,l,l',$\ell$,$\ell'$,J,q,s][t,y,x][$f^{J,\ell',0}_{l,\ell,l',\ell'}(t,y,x)$]} \, .
\end{align*}
\end{small}

    \item From polynomial in $(y,x)$, to Gelfand-Tsetlin basis\footnote{Theoretical background for the map and the subsequent normalisation is given in Appendix \ref{From polynomials to the Gelfand-Tsetlin basis}}, $$\texttt{toGTbasis\$MSTMST[d,l,l',$\ell$,$\ell'$][t,y,x][f]}\ .$$
    \item Normalisation of matrix elements, $\texttt{normaliseMSTMSTinGT[d,l,l',$\ell$,$\ell'$,J,q,s][$\theta$][f]}$. Notice that the result is a $(\text{min}(l,l')-\ell' +1)\times(\ell+1)$ matrix.
\end{itemize}

\subsection{A comment about internal weight-shifting}
Rarely, there may be cases where the output of one our functions \eqref{GenerateFunctions-nb} is zero. This is of course wrong, and it is due to the nature of the internal weight-shifting procedure. Namely, when we do the tensor decomposition by means of multiplying a function with the zonal spherical function, e.g.
\begin{equation*}
  (J,q) \otimes (k) = (J+k,q) \oplus \dots \,,
\end{equation*}
on the right hand side their may be different representations with the same value of the quadratic Casimir. This creates a problem when we need to project one of these representations out, since all the others are automatically also projected out, giving zero as the final result. Hence, the problem is that the Laplacian cannot distinguish between two functions with different internal representations with the same Casimir value. This is exactly what is happening in the STT-STT case in $d=5$, where $f^{J,q}_{l,l'}(t,y)$ and $f^{J,-q}_{l,l'}(t,y)$ have the same eigenvalue. As in that case, a solution to this problem would be to use another Casimir which can distinguish those representations (or alternatively, find another tensor decomposition which does not exhibit the same problem).

\section{Discrete symmetries}
\label{appendix:discrete symmetry}

In Section \ref{From partial waves to matrix elements} we provided a construction of three-point tensor structures. Similarly, the space of four-point tensor structures admits a simple group-theoretic characterisation, \cite{Kravchuk:2016qvl}. For external particles of spin $\pi_1,\dots,\pi_4$ and the exchanged one of spin $\pi$
\begin{equation}\label{spaces-3pt-4pt-structures}
    T_3(\pi_1,\pi_2,\pi) = (\pi_1\otimes\pi_2\otimes\pi^\ast)^{SO(d-2)}, \quad T_4(\pi_1,\dots,\pi_4) = (\pi_1\otimes\dots\otimes\pi_4)^{SO(d-3)}\ .
\end{equation}
Here, the notation $V^G$ stands for the space of $G$-invariants in $V$. Partial waves \eqref{partial-waves-shadow-integral} automatically satisfy kinematical constraints of a four-point function, i.e. can be expanded in four-point tensor structures. In will become clear in the next section how the $SO(d-3)$ invariance in manifest in the formula \eqref{matrix-elements-from-pw} for partial waves. Furthermore, we will make a particular choice of basis for $T_4(\pi_1,\dots,\pi_4)$ for any arbitrary external spins. By that point, we will have taken account of \eqref{spaces-3pt-4pt-structures}. Additional kinematic constraints on the scattering amplitude are present in the case when some of the particles are identical or symmetry under parity is imposed. The resulting partial waves are constructed out of non-symmetric matrix elements \eqref{matrix-elements-from-pw} by means of some simple operators. Since this part of the theory does not differ from other accounts, we only illustrate the salient points, following \cite{Kravchuk:2016qvl,Chowdhury:2019kaq,Chakraborty:2020rxf}. 

\paragraph{Permutations: four-point functions} Let $S_4$ be the permutation group of the four particles and consider the normal subgroup $H \cong \mathbb{Z}_2 \times \mathbb{Z}_2$ generated by double transpositions\footnote{We denote permutations using their presentation in terms of disjoint cycles.}
\begin{equation}
    H = \langle(12)(34),(13)(24)\rangle\ .
\end{equation}
This subgroup preserves the Mandelstam variables. For identical particles, the space of four-point tensor structures is
\begin{equation}\label{4pt-permutation-invariants}
    \left((\pi_1\otimes\dots\otimes\pi_1)^{\mathbb{Z}_2\times\mathbb{Z}_2}\right)^{SO(d-3)} \cong \left(\pi_1^{\otimes4} \ominus 3(S^2\pi_1\otimes\Lambda^2\pi_1)\right)^{SO(d-3)}\, ,
\end{equation}
where $S^2$ and $\Lambda^2$ denote the symmetric and antisymmetric square, respectively.
Any four-point amplitude may be expanded in these structures and is further constrained by the residual symmetry
\begin{equation}
    S_3 = \frac{S_4}{\mathbb{Z}_2 \times \mathbb{Z}_2}\ .
\end{equation}
The residual symmetry group acts non-trivially both on the Mandelstam variables and on four-point tensor structures.

\paragraph{Three-point functions} The simplest of permutation symmetries to impose for three-point functions is under the transformation $(12)(34)$. To this end, we consider two subspaces of three-point structures
\begin{equation}\label{odd-even-3pt-structures}
     T_3^{even}(\pi_1,\pi_1,\pi) = (S^2\pi_1\otimes\pi^\ast)^{SO(d-2)},\quad T_3^{odd}(\pi_1,\pi_1,\pi) = (\Lambda^2\pi_1\otimes\pi^\ast)^{SO(d-2)}\ .
\end{equation}
In cases where $\pi_1^{\otimes2}$ is multiplicity-free over $SO(d-2)$, our choice of three-point structures made in Section \ref{Single particle states} is compatible with the above split. Assume that $\pi_1^{\otimes2}$ is not multiplicity-free and that the irreducible representation $\pi_{12}^{(1)}$ appears with multiplicity $k$. One can always choose the labelling of different copies $\pi_{12}^{(1)},\dots,\pi_{12}^{(k)}$ such that each of them is a subspace of either $S^2\pi_1$ or $\Lambda^2\pi_1$. With such labelling, the choice of three-point structures is always compatible with \eqref{odd-even-3pt-structures}.

\paragraph{Example} Before imposing permutation symmetry, there are ten four-point structures for scattering of photons in $d\geq5$. Indeed, two-particle spins read
\begin{equation}\label{product-of-two-photons}
    (1) \otimes (1) = (2) \oplus (0) \oplus (1,1) = (2) \oplus 2(1) \oplus 2(0) \oplus (1,1)\ .
\end{equation}
In the last step, we have restricted the $SO(d-2)$ representation to $SO(d-3)$. Using the fact that a singlet only appears (with multiplicity one) when an irreducible representation is tensored with itself, the number of four-point structures is equal to the sum of squares of multiplicities in the last expression, i.e. 10. To account for permutations, use
\begin{equation}\label{symmetric-products}
    S^2\pi_1 = (0) \oplus (2),\qquad \Lambda^2\pi_1 = (1,1)\ .
\end{equation}
Counting $SO(d-3)$-invariants is the same as counting the number of STT components of the expression inside the last bracket in \eqref{4pt-permutation-invariants}. In the case at hand, the symmetric traceless part is found to be $(4)\oplus3(2)\oplus3(0)$. Thus, there are seven four-point structures. In $d=5$, \eqref{product-of-two-photons} and \eqref{symmetric-products} get modified, but the number of tensor structures is again seven.

\paragraph{Parity} So far, we have focused on the connected component of the identity in the Lorentz group $O(d-1,1)$. The full Lorentz group further includes parity and time-reversal symmetries which we now briefly discuss. The subgroup of $O(d-1,1)$ that stabilises a pair of massless momenta is $O(d-2)$. Therefore, the $SO(d-2)$-invariants are further distinguished as either scalars of pseudo-scalars of $O(d-2)$. If parity is a symmetry of the theory, only the former are allowed. Also, particles in a theory with $O(d-1,1)$ symmetry transform in representations of $O(d-1)$. These carry an additional label compared to $SO(d-1)$. Namely, representations of $O(2r+1)$ are labelled by $(l_1,\dots,l_r)^{\pm}$, where $l_1,\dots,l_r$ are the usual Gelfand-Tsetlin labels for a representation of $SO(2r+1)$ and irreducible representations of $O(2r)$ are labelled in either of the following ways
\begin{equation}
    (l_1,\dots,l_r\neq0), \quad\text{or}\quad (l_1,\dots,l_r=0)^\pm\ .
\end{equation}
The restrictions between consecutive orthogonal groups depend on additional labels. In the end, we get further restrictions on quantum numbers of intermediate particles.

\paragraph{Example} Photons in $d=5$ are characterised by the representation $(1)^+$ of $O(3)$. The two-particle spins thus transform in
\begin{equation}
    (1)^+ \otimes (1)^+ = (2)^+ \oplus (1)^+ \oplus (0)^+\ .
\end{equation}
Without imposing parity, possible exchanged particles have spins $(J)$, $(J,\pm1)$ and $(J,\pm2)$. The $SO(4)\downarrow SO(3)$ branching rules imply that the number of three-point structures for these exchanges are (for sufficiently large $J$) 3, 2 and 1, respectively. Taking parity into account, the exchanged particles can have labels $(J)^\pm$ or $(J,q)$, with $q=1,2$. We focus on the cases $(J)^+$ and $(J,q)$. Written as $SO(4)$ representations, these are $(J)$ and $(J,q)\oplus(J,-q)$. 

Putting both parity and permutation symmetries together we arrive at the following. The exchanged $SO(4)$ representations and the numbers of three-point structures are
\begin{align*}
    & \text{multiplicity two}: \quad (2k) \to \{(0),(2)\}\,,\\
    & \text{multiplicity one}: \quad (0)\to \{(0)\}, \quad (2k+1)\to\{(1)\}\,,\\
    & \hspace{3cm} (2k,\pm1)\to \{(1)\} , \quad (2k+1,\pm1) \to \{(2)\}, \quad (2k,\pm2) \to \{(2)\}\ .
\end{align*}
Next to each exchanged representation, we have written the corresponding three-point structures. Partial waves in terms of $SO(4)$ matrix elements read
\begin{equation}
    \Pi^{(0)\mu}{}_{(0)\nu},\quad \Pi^{(1)\mu}{}_{(1)\nu}, \quad \Pi^{(0)\mu}{}_{(2)\nu} + \Pi^{(2)\mu}{}_{(0)\nu}, \quad \Pi^{(2)\mu}{}_{(2)\nu}\ .
\end{equation}
Here $\Pi$ denotes either the matrix element of the irreducible representation $(J)$ or the sum of matrix elements of the two representation $(J,q)$ and $(J,-q)$.

\bibliographystyle{JHEP}
\bibliography{bibliography}

\end{document}